\documentclass[useAMS,usenatbib, usegraphicx]{mn2e}
\usepackage{aas_macros}
\usepackage[dvips]{graphicx} 

\title[The impact of a non-universal IMF on the SFHs of ETGs]{The impact of a non-universal Initial Mass Function on the Star Formation Histories of Early-Type Galaxies}
\author[A. Ferr\'e-Mateu]{A.Ferr\'e-Mateu$^{1,2}$\thanks{E-mail:
aferre@iac.es (AFM)}, A. Vazdekis$^{1,2}$ and I. G. de la Rosa$^{1,2}$\\
$^{1}$Instituto de Astrof\'isica de Canarias, V\'ia L\'actea s/n, La Laguna, 38200, Spain\\
$^{2}$Departamento de Astrof\'isica, Universidad de La Laguna, Spain}
\begin{document}

\date{Accepted for publication in MNRAS.}

\pagerange{\pageref{firstpage}--\pageref{lastpage}} \pubyear{2011}

\maketitle

\label{firstpage}

\begin{abstract}
{Recent results on the non-universality of the Initial Mass Function (IMF) have shown strong evidence of IMF variations with galaxy velocity dispersion, with a corresponding impact on other stellar population parameters, line indices and colours. Using a set of stellar population models with varying IMF slopes, we provide additional caveats on the assumption of a universal IMF. The present study shows that the derived star formation histories of early-type galaxies vary significantly with the IMF slope. For instance, a steepening in the slope of a single power-law IMF decreases substantially, by a factor of up to four, the contribution of the old stellar populations to the total light/mass. This trend is milder for a segmented-like IMF shape, where the contribution of the very low mass stars is decreased. It is also shown that, by tuning each IMF slope to its prescribed value according to each galaxy velocity dispersion, a sample of early-type galaxies covering a range of masses yield comparable star formation histories. On the one hand, a small contribution from relatively young stellar populations appear in the star formation histories of most massive elliptical galaxies when adopting a steep IMF. On the other hand, we find that low mass early-type galaxies that look like genuinely young objects with a standard IMF (i.e. ``\textit{baby elliptical galaxies}``), turn out to be older when a slightly flatter IMF is employed. In summary, the use of a non-universal IMF, tuned according to the velocity dispersion of the galaxy, seems to provide more consistent results.}

\end{abstract}


\section{Introduction}
The Initial Mass Function (IMF), which is the distribution of stellar masses in a single population at the time of birth, has been usually assumed to be universal. This means that we consider the same IMF at all cosmic times, no matter the environment or the mass of the galaxy. This universal assumption was adopted mainly due to the difficulties in obtaining direct constrains. So far, the IMF can only be directly constrained from our own Milky Way by counting individual stars \citep{Salpeter1955}. Many works have supported this universality, warning us about the consequences of freely varying it (see e.g. \citealt{Gilmore2001}, \citealt{Bastian2010}, \citealt{Greggio2012}, \citealt{Narayanan2012}). For example, an IMF with a steeper slope than Salpeter would probably fit better the near-infrared Ca{\sc II} triplet, but it would also produce too red V-K colours and too high M/L (e.g. \citealt{Cenarro2003}).\\
Nonetheless, scenarios with a non-universal IMF have been discussed at length, starting with \citet{Schmidt1963} and followed by \citet{Faber1979}, \citet{Tinsley1980}, \citet{Worthey1992}, \citet{Elbaz1995}, \citet{Vazdekis1996} or \citet{Vazdekis1997}, among others. Recently, abundant observational evidence from a variety of independent techniques have pointed out that a non-universal IMF might be in fact the real case (\citealt{Tortora2012} and references therein). From the measurements of spectral features sensitive to the slope of the IMF in massive giant elliptical galaxies (e.g. Ca{\sc II} triplet, the Wing-Ford band and the NaI8200), it has been discovered the existence of a larger population of low-mass stars than predicted by a Kroupa or even a Salpeter IMF (e.g. \citealt{Vazdekis2003}, \citealt{Cenarro2003}, \citealt{Falcon-Barroso2003}, \citealt{vandokkum2010}, \citealt{Conroy2012a} (CvD12 hereafter), \citealt{Spiniello2012},  \citealt{vanDokkum2012}, \citealt{Conroy2012b}). Colours are also highly sensitive to IMF variations, in the sense that redder galaxies have bottom-heavier IMFs, i.e. steeper slopes (see e.g. \citealt{Dutton2012a}, \citealt{Pforr2012}, \citealt{Ricciardelli2012}, \citealt{Vazdekis2012}). In addition, recent works have added strong evidence on the non-universality of the IMF by constraining the dark matter fraction in galaxies, with gravitational lensing and dynamical studies, finding an IMF dependence with galaxy mass (e.g: \citealt{Grillo2009}, \citealt{Auger2010}, \citealt{Napolitano2010}, \citealt{Treu2010}, \citealt{Barnabe2011}, \citealt{Thomas2011}, \citealt{Cappellari2012b}, \citealt{Dutton2012c}, \citealt{Sonnenfeld2012} or \citealt{Tortora2012}). Moreover, in \citet{Ferreras2013} (F13 hereafter) the authors have shown a tight correlation between the velocity dispersion of early-type galaxies and the IMF slope, in the sense that higher velocity dispersion galaxies need steeper IMF slopes, while the low mass ones need flatter slopes. \citet{Cappellari2012a} have shown a similar result, where the use of different IMF slopes varies the amount of dark matter and implies a change in the dynamical masses by a factor of three in early-type galaxies.\\ 
As the IMF is a parameter that clearly influences most of the stellar population properties, its determination is crucial to understand the formation and evolution of galaxies. Most works in the field of the stellar populations are based on its assumed universality (e.g. \citealt{Bender1992}, \citealt{Kuntschner1998}, \citealt{Jorgensen1999}, \citealt{Poggianti2001}, \citealt{Terlevich2001}, \citealt{Blanton2003}, \citealt{Kauffmann2003}, \citealt{Gallazzi2006}, among many others). However, a number  of still open questions could be addressed by changing the IMF according to the recent claims in this direction.\\
\\
The work here conducted has been devised as a ''user-guide`` for observers, who usually apply their analysis tools using models assuming a standard, universal IMF such as Kroupa \citep{Kroupa2001}, Chabrier \citep{Chabrier2003} or Salpeter \citep{Salpeter1955}. Rather than prescribing a particular IMF, it is shown here that more consistent results can be obtained when the IMF slopes are tuned according to the velocity dispersion of the galaxies. Following the trend in F13, we will use throughout this work the notation $\mu(\sigma)$ for these $\sigma$-dependent IMF slopes. We analyze the behavior of the Star Formation History (SFH) and the derived stellar mass (M$_{star}$) of Early-Type Galaxies (ETGs) when models with different IMF slopes and shapes are used.\\
Section 2 describes the high quality long-slit data sample used to derive accurate SFHs for galaxies covering a wide range of masses. In Section 3 we analyze the impact of varying the IMF slope on various galaxy properties such as the M/L ratio, the SFH, the derived ages or the stellar masses. Section 4 is dedicated to an illustrative case and our conclusions and summary are found in Section 5. We have adopted a concordance cosmology with $\Omega_{m}$=\,0.3, $\Omega_{\Lambda}$=\,0.7 and H$_{0}$=\,70\,km\,s$^{-1}$ Mpc$^{-1}$.

\section{The Data}
Spectra from two ETGs samples are used for the present study. The main sample, with high-quality long-slit spectra make the bulk of the work, while a second sample with fiber spectra from the SDSS (\textit{Sloan Digital Sky Survey}) is used for statistical purposes.\\
The first sample, with S/N$\geq$\,40$\,\rm\AA{}^{-1}$, was constructed with spectra from \citet{Sanchez-Blazquez2006a} (PSB06), \citet{Yamada2006} (Y06) and \citet{Ferre-Mateu2012} (AFM12). The reader is referred to these papers for a complete description of the data, their quality and reduction process. The first two references contain ETGs covering a range of masses ($\sigma$=\,[50-300]\,$\rm{km\,s^{-1}}$), while the third one includes massive compact galaxies that are found in the Local Universe. The latter are a family of unique galaxies found in \citet{Trujillo2009} and characterized in AFM12 and \citet{Trujillo2012}. All these galaxies allow us to cover an ample range of properties, like masses, luminosities, sizes, stellar populations, etc. Therefore, this sample is ideal to detect the effects of IMF variations. In particular, the Y06 sample covers the colour-magnitude diagram of the Virgo Cluster. At the very high mass regime ($\sigma\sim$\,300\,$\rm{km\,s^{-1}}$) galaxies with the steepest IMF are located, according to F13 and CvD12. The local compact massive galaxies were selected because they have revealed unusually large fractions of young components on their derived SFHs. The massive ellipticals were selected to have similar velocity dispersions to the compact galaxies, albeit larger radii, to compare their mass estimates ($\sigma$=\,200\,$\rm{km\,s^{-1}}$). The low-mass ellipticals are useful probes to study the SFH differences, as a fraction of them show young ages, comparable to those in the massive compacts. Table 1 lists all the galaxies in the high-quality spectra sample with their main properties. Despite their different origins, they were all treated and analyzed in a similar way. \\
The sample of ETGs from the SPIDER project, with SDSS spectra of 40,000 objects (\citealt{LaBarbera2010}, LB10 hereafter), has been used as the benchmark for sanity checks performed to test the results obtained. It was also used to find rather extreme, but illustrative, cases in Section 4.

\begin{table*}
\centering
\caption{Galaxy sample of high-quality long-slit spectra}           
\label{table:1}     
\begin{tabular}{c| c c c c c}   
\hline\hline      
Galaxy ID &  Source & environment & type & z & $\sigma$($\rm{km\,s^{-1}}$) \\    
\hline  
 NYU\,54829  & AFM12 & Field & MC & 0.085 & 137 \\      
 NYU\,321479 & AFM12 & Field & MC & 0.128 & 221 \\     
 NYU\,685469 & AFM12 & Field & MC & 0.149 & 204 \\    
 NYU\,796740 & AFM12 & Field & MC & 0.182 & 203 \\     
 NYU\,890167 & AFM12 & Field & MC & 0.143 & 233 \\      
 NYU\,896687 & AFM12 & Field & MC & 0.130 & 223 \\       
 NYU\,2434587& AFM12 & Field & MC & 0.172 & 206 \\
 NGC\,4365 & Y06   & Virgo & HME & 0.0041 & 265  \\
 NGC\,4472 & Y06   & Virgo & HME & 0.0033 & 300 \\
 NGC\,2329 & PSB06 & Coma  & ME  & 0.0192 & 225 \\ 
 NGC\,4473 & Y06   & Virgo & ME  & 0.0074 & 193 \\
 NGC\,4621 & PSB06 & Virgo & ME  & 0.0013 & 230 \\ 
 NGC\,4697 & Y06   & Virgo & ME  & 0.0041 & 180 \\ 
 NGC\,5812 & PSB06 & Field & ME  & 0.0065 & 215 \\ 
 NGC\,4239 & Y06   & Virgo & LME & 0.0031 & 63  \\  
 NGC\,4339 & Y06   & Virgo & LME & 0.0043 & 114 \\
 NGC\,4387 & Y06   & Virgo & LME & 0.0015 & 98  \\
 NGC\,4458 & Y06   & Virgo & LME & 0.0210 & 105 \\ 
 NGC\,4464 & Y06   & Virgo & LME & 0.0041 & 135 \\
 NGC\,4467 & Y06   & Virgo & LME & 0.0047 & 69  \\ 
 NGC\,4489 & Y06   & Virgo & LME & 0.0032 & 52  \\ 
 NGC\,4551 & Y06   & Virgo & LME & 0.0039 & 105 \\  
\hline                                  
\end{tabular}\\
{Main galaxy properties: (1) galaxy ID; (2) publication source: PSB06 \citep{Sanchez-Blazquez2006a}, Y06 \citep{Yamada2006} and AFM12 \citep{Ferre-Mateu2012}; (3) environment of the galaxy; (4) galaxy type: MC (Massive Compact), HME (High Mass Elliptical), ME (Massive Elliptical) or LME (Low-Mass Elliptical);  (5) redshift; and (6) central velocity dispersion. Values are taken from the publication source or from the LEDA database (http://leda.univ-lyon1.fr/) when not available. }
\end{table*}

\section{Galaxy properties with varying IMF}
A study on the quantitative impact of varying the IMF shape and slope on the stellar population properties is presented in this section.
A first consideration should be done regarding the shape of the IMF assumed for the models. Historically, two alternative IMFs have been considered: a single-power law over all stellar masses, characterized by its logarithmic slope $\mu$ \citep{Salpeter1955} and a multi-segmented IMF \citep{Kroupa2001}. These two IMFs will be regarded as standard through this work. For our illustrative purposes, the other popular choice, the log-normal IMF \citep{Chabrier2003} is very similar to that of Kroupa and will not be considered here.

\subsection{Mass-to-light ratios}
We can study the dependence of the mass-to-light ratio (M/L) with the IMF using stellar population synthesis models. We have selected three representative ages (1, 2 and 10\,Gyr) and three metallicities ([Z/H]= $-$0.40, 0.00 and 0.22) from the extended MILES single-burst models (MIUSCAT; \citealt{Vazdekis2012}, V12 hereafter).  We have selected six different IMF slopes: $\mu=\,$[0.3, 0.8, 1.0, 1.3, 1.8 and 2.3] both for a single- and segmented-power law (unimodal and bimodal following the notation of \citet{Vazdekis1996}). As a reference, $\mu$\,=\,1.3 corresponds to a Salpeter IMF slope with a unimodal shape. Instead, a bimodal IMF with the same slope resembles closely to a Kroupa IMF. In addition, a unimodal IMF with slope $\mu\sim$\,1.0 would render similar results in the studied properties to the Chabrier IMF.\\ 
Figure 1 shows the dependence of the M/L with the slope of the IMF for stellar populations of different ages with both shapes. It can be seen that the M/L from the stellar population models presents a minimum. For the M/L calculation both the alive stars contributing to the light and the stellar remnants are taken into account. Whereas the steepest IMFs emphasize the contribution of long-lived low-mass stars, the flatter IMFs favour the number of remnants resulting from short-lived massive stars. Note that if the stellar remnants were not included, the resulting M/L would not show a minimum when varying the IMF slope (see the predictions shown in the MILES webpage\footnote{http://miles.iac.es/}). The location of this minimum depends on the age of the SSP: younger ages reach their minimum at flatter slopes. In the unimodal case, the minimum is found at $\mu$\,$\sim$\,0.8 for young populations, whereas for it is located at $\mu$\,$\sim$\,1.0 for the old ones. In the bimodal case the minimum is seen at $\mu$\,$\sim$\,1.0 (1.3) for the young (old) populations. It is worth noticing the strong variation of the M/L with the IMF slope. For the unimodal shape, the M/L value increases strongly for both the flattest and the steepest IMFs. For the bimodal IMF a significantly milder variation is seen, particularly for the oldest stellar populations (see also F13). The choice of one shape over the other will be affecting in a different way the derived stellar populations properties, such as the M/L. The plot also shows a trend with the broad band filters. At a fixed IMF slope, lower M/L values are obtained for the redder filters.                                    

\begin{figure*}
\includegraphics[scale=0.65]{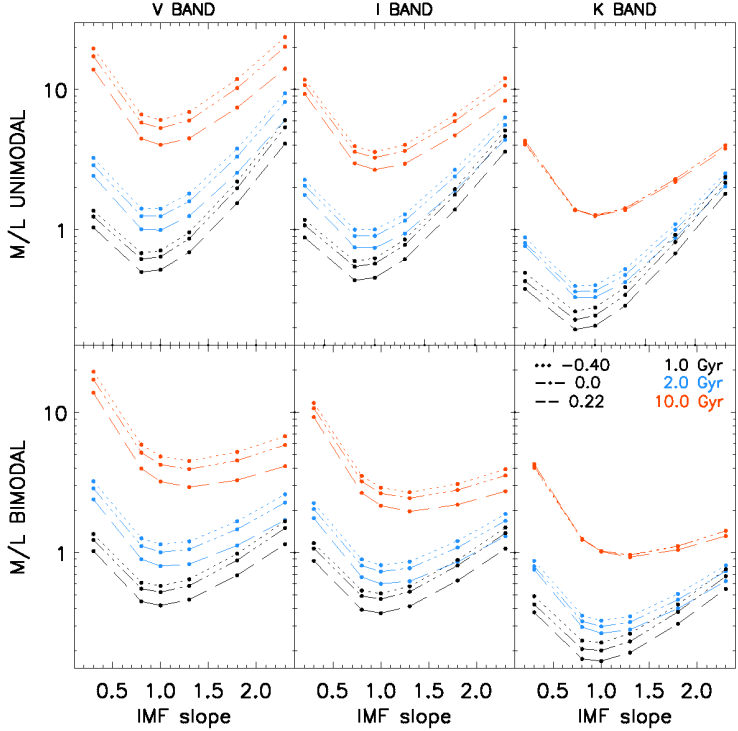}
\label{Fig.1}
\caption{The M/L ratio in several broad-band filters is plotted against the IMF slope for models of 1\,Gyr (black), 2\,Gyr (blue) and 10\,Gyr (red) and for different metallicities ([Z/H]=\,$-$0.40, dotted line; [Z/H]=\,0.00, dashed-dotted line and [Z/H]=\,0.22, dashed line). Upper rows correspond to models with a unimodal IMF and lower ones to models with a bimodal IMF. A minimum on the M/L is seen around $\mu\,\sim$\,0.8-1.0 (unimodal) or $\mu\,\sim$\,1.0-1.3(bimodal), depending on the SSP age.}
\end{figure*} 

\subsection{Star Formation Histories}
To obtain the SFHs of our high-quality galaxy sample we apply a full-spectrum-fitting approach. We use the code {\tt STARLIGHT} \citep{CidFernandes2005} with the MIUSCAT SSP SEDs. This code fits a set of SSP models to the observed spectra and selects the best population mixture. The SFH is derived for each galaxy using different base models with all the IMF slopes selected above, both for the unimodal and the bimodal cases. \\
The high quality of our spectra allows us to perform an accurate fitting. We are aware that our results may be biased due to the relatively short spectral range of the spectra (3800-5300$\,\rm\AA{}$). Although this range encompasses most of the commonly used absorption line features, the reported IMF-sensitive indices are located at redder wavelengths. However, the blue part of the spectrum also shows some sensitivity to the IMF, as shown in figure 9 of V12. To check the robustness of our results, we have performed various tests to take into account different sources that might change our results (see Appendix A). We first masked different features in the spectra, finding that the variations on the mean derived ages are negligible (smaller than 2$\%$). We also compared the results from selecting a short \textit{versus} a longer spectral range, which may account for variations in the derived luminosity(mass)-weighted ages of the order of a 8$\%$ (12$\%$). Finally, due to the sensitivity of some colours to the IMF variations (see e.g. \citealt{Ricciardelli2012}), we have compared the synthetic colours obtained from our derived SFHs to those from the SDSS broadband photometry. Note that the latter extend to the UV and the IR, i.e. sampling the different populations in the galaxy. We find a good agreement between both. These tests confirm the robustness of the results and trends obtained in Section 3. From now on, all the SFHs presented here are derived using the spectral range of our high quality long-slit sample.\\
\\
For illustration, Figure 2 shows the SFHs of four galaxies, colour-coded by galaxy family: the very massive galaxy NGC\,4472 (black), the massive elliptical NGC\,4473 (green), the massive compact NYU\,2434587 (purple), and the low-mass galaxy NGC\,4458 (yellow). The panels show the derived SFHs with base models where the IMF slope steepens from left to right. In the upper rows we show the results for the unimodal case and in the lower ones for the bimodal one. The dashed vertical line shows the mean mass-weighted age for each varying slope, while the solid line is the mean value derived from the standard $\mu$=\,1.3 slope. The SFHs for the other galaxies are shown in Appendix B.\\ 
Some interesting conclusions are drawn from Fig. 2. First, we see noticeable differences in the derived SFHs for these types of galaxies: while massive ellipticals are mainly old, both in light and mass, massive compacts show an unusually large fraction of young stellar populations ($\leq$\,2\,Gyr), both light- and mass-weighted. We refer the reader to AFM12 for an extensive discussion on this issue. Second, we see a general trend of decreasing the fraction of old stellar populations with increasing IMF slope. This trend is shown in Fig. 3 (upper panel), where we indicate how the fraction of stellar populations larger than 5\,Gyr evolves with the IMF slope.\\
This trend is seen in all galaxies and for both IMF shapes studied here, though it is emphasized for the unimodal case. For various massive galaxies a stellar population with age $\sim$5\,Gyr emerges when adopting an IMF steeper than Salpeter, leading to a more complex SFH. The impact is significantly milder for the bimodal case. Note however, that according to F13 the best fits to the various IMF-sensitive features for the most massive galaxies require an IMF slope around 2 and 3 for the unimodal and bimodal shape, respectively.\\
This rejuvenating effect implies a decrease in the mean age of the galaxy, as shown in the lower panel of Fig. 3, where the derived mean mass-weighted ages are shown with respect to the IMF slope. It is seen that in the unimodal case, the derived ages might decrease by a factor of as much as 3 from $\mu$=\,0.8 to $\mu$=\,2.3, while for the bimodal case is 1.5.\\ 
\\ 
As stated before, recent results have claimed that more massive galaxies demand steeper IMF slopes than Salpeter (e.g. \citealt{Cappellari2012a}, \citealt{vanDokkum2012}, CvD12, \citealt{Spiniello2012}), including a systematic correlation with the $\mu$-$\sigma$ found by F13. These papers pose an upper limit for the steepest slope around $\mu\sim$2 for a single power-law IMF
shape. Therefore, for our illustrative purpose we will only discuss the unimodal case in this part of the work. Which SFHs will be derived for our objects if we follow the $\mu$-$\sigma$ relation? We find that most massive galaxies for which we adopt $\mu(\sigma)\sim$2 are very old. However, for most cases, a non-negligible fraction of relatively younger stellar components tends to appear
(up to 10$\%$). This also applies to the massive galaxies, for which a Salpeter or slightly steeper IMF is demanded. On the contrary, the contribution from old stellar populations increases for the low-mass ETGs if adopting flatter slopes than Salpeter ($\mu\sim$0.8-1.0).\\ 
In summary, our results suggest that the SFHs of ETGs of all masses tend to converge to a common pattern, when the prescribed IMF slopes, according to the galaxy velocity dispersion, are used. This common SFH pattern involves a varying amount of recent residual star formation. The SFHs of ETGs have always been a challenge in modern astronomy, as they represent a direct way to constrain competing galaxy formation models. Although it is well established that the bulk of the stars in ETGs formed at high redshift (z$\sim$2), a small ''frosting`` of young stars forms on top of it at more recent epochs (e.g: \citealt{Trager2000}). Observations in the UV are particularly useful to trace these recent bursts, as the rest-frame UV is mostly sensitive to stars less than $\sim$1\,Gyr old (e.g: \citealt{Ellis2001}, \citealt{Kaban2005}, \citealt{Kaviraj2005}, \citealt{Yi2005}, \citealt{Ferreras2006}). All these works show that there is indeed this residual
star formation, which evolves with redshift (\citealt{Fukugita2004}, \citealt{Kaviraj2008}). It can be up to $\sim$10$\%$ in the local Universe (e.g \citealt{Kaviraj2007}, \citealt{Nolan2007}), in agreement with our results. \\
Nonetheless, for the compact massive galaxies with an unusually large contribution of young stellar populations, as reported in AFM12, the derived SFHs do not change substantially if the IMF is varied according to the $\mu$-$\sigma$ relation. This also applies when adopting the more conservative bimodal IMF shape. Therefore, these galaxies further show their unique nature in the present analysis.\\ 
Table 2 summarizes all the mass-weighted properties derived from the SFHs that are obtained when adopting for each galaxy their $\mu(\sigma)$. The other derived quantities for both the two IMF shapes adopted in this work can be found in Table B1 and B2 in Appendix B. 

\begin{figure*}
\includegraphics[scale=0.5]{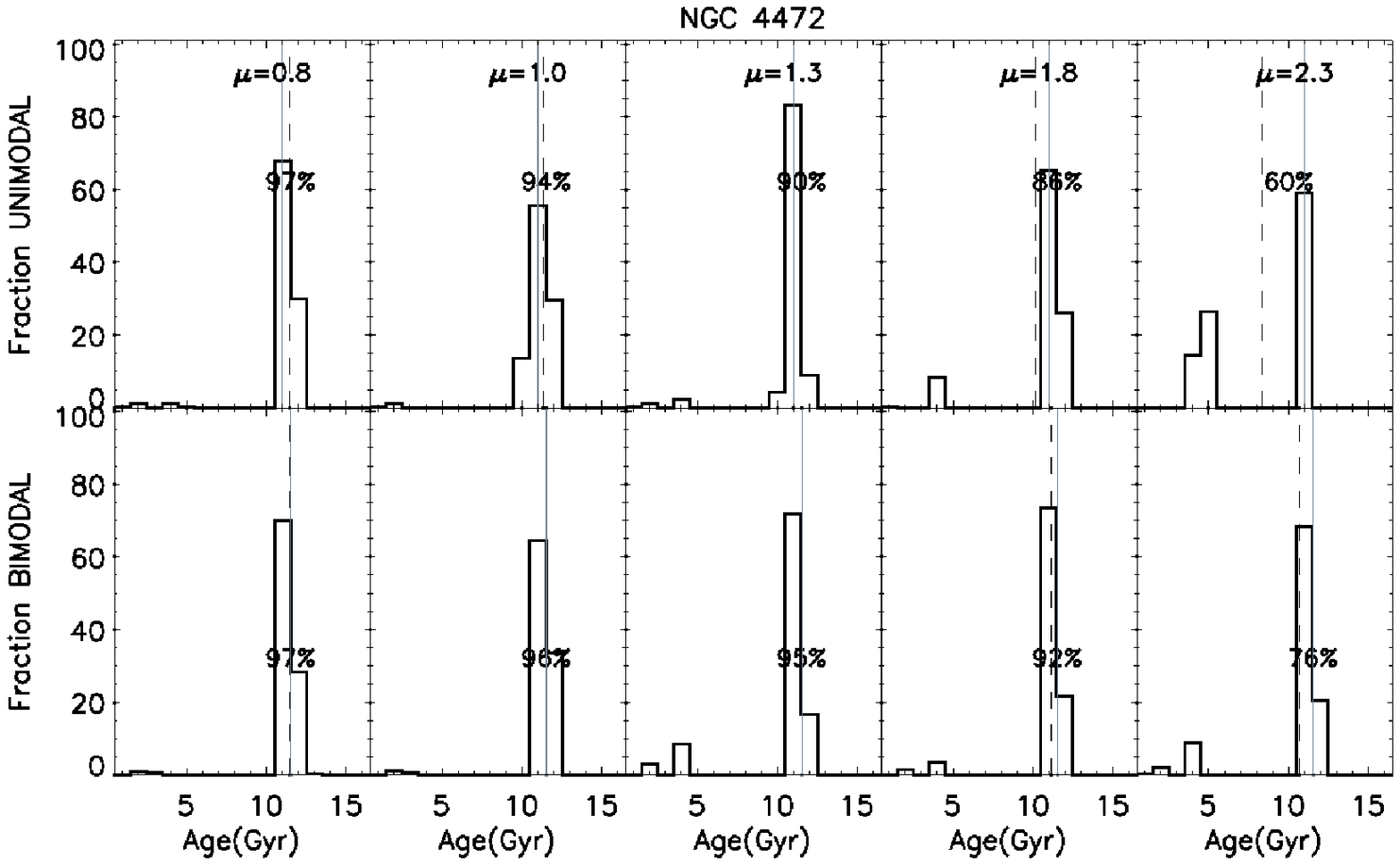}
\includegraphics[scale=0.5]{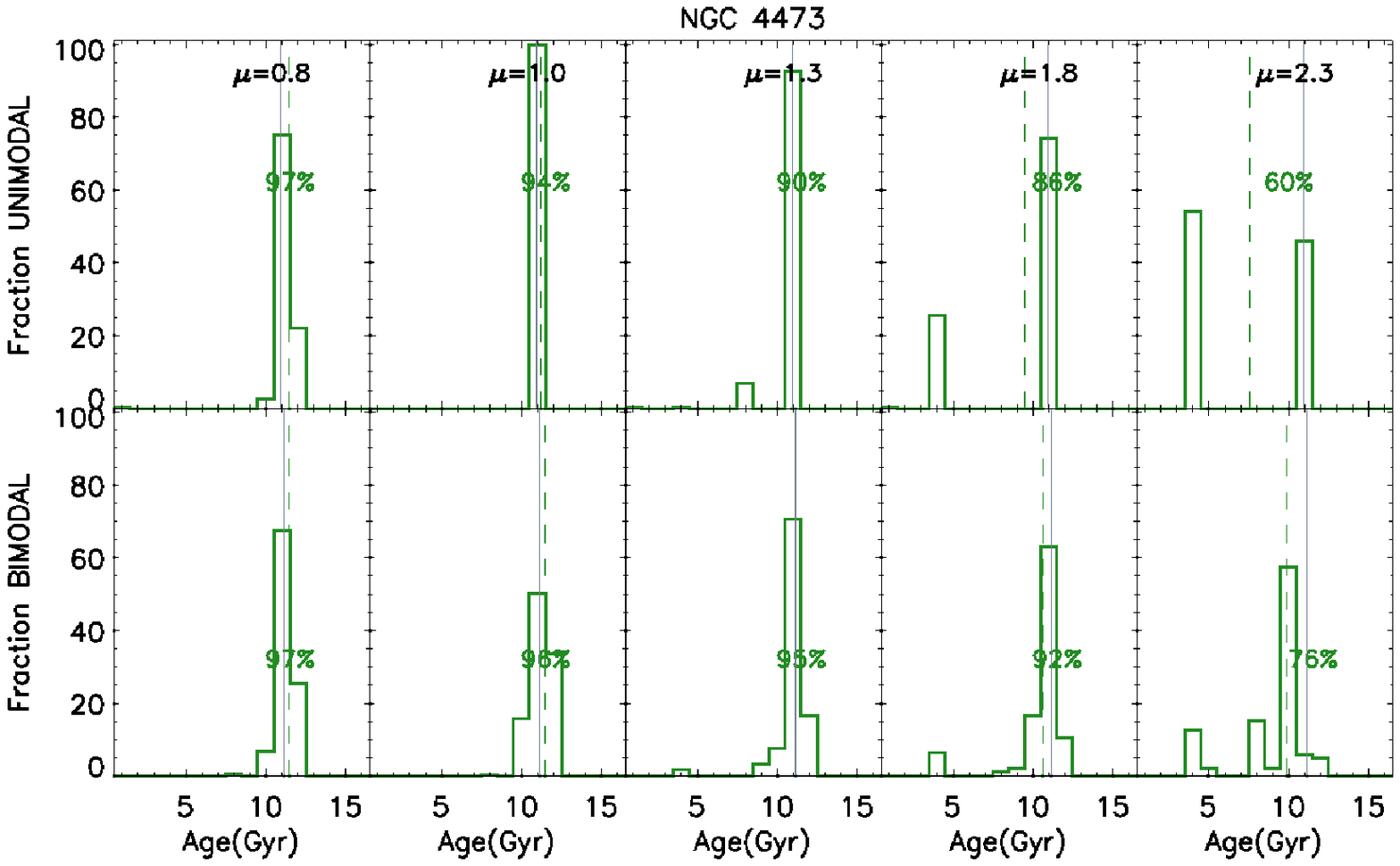}\\
\includegraphics[scale=0.5]{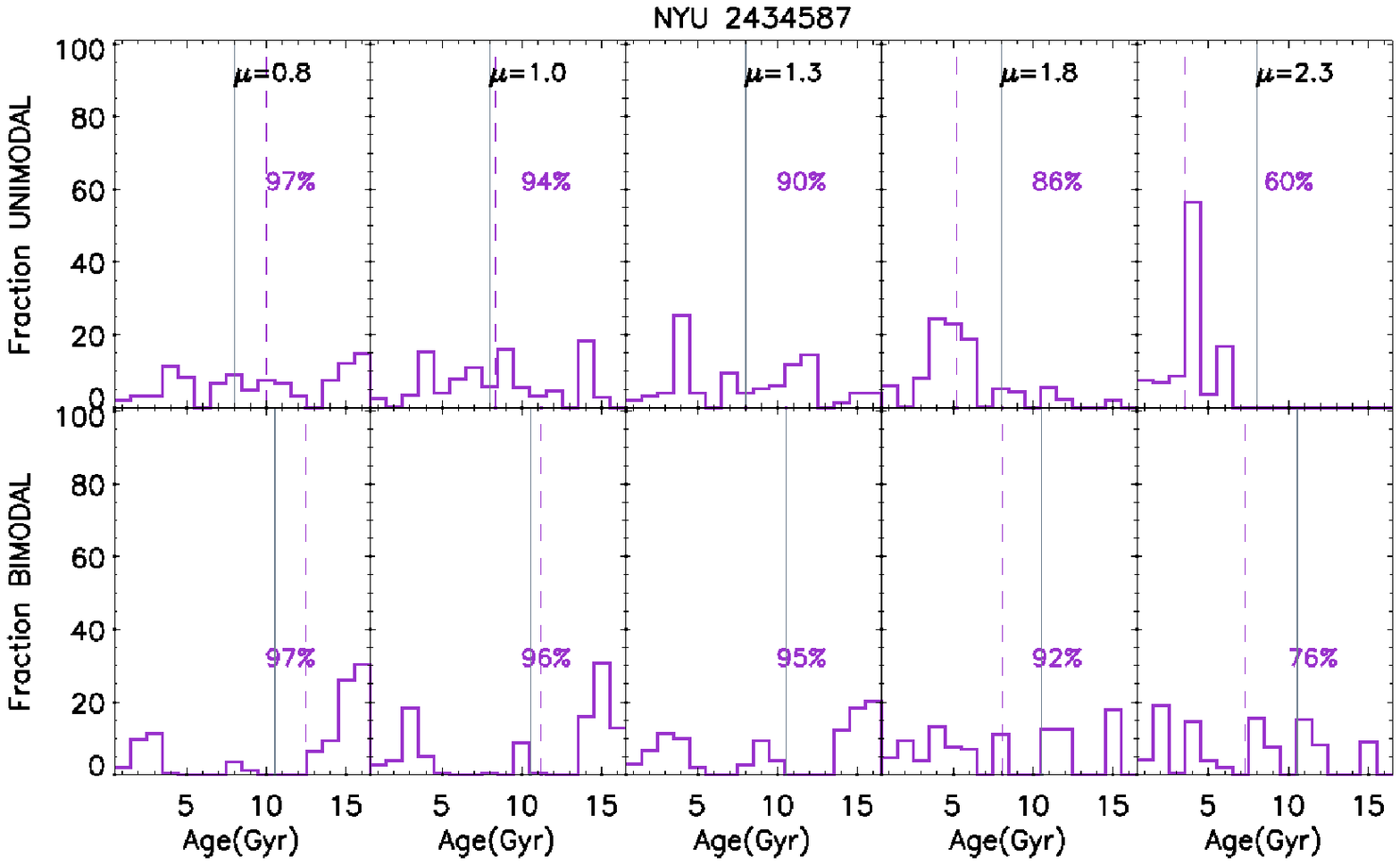}
\includegraphics[scale=0.5]{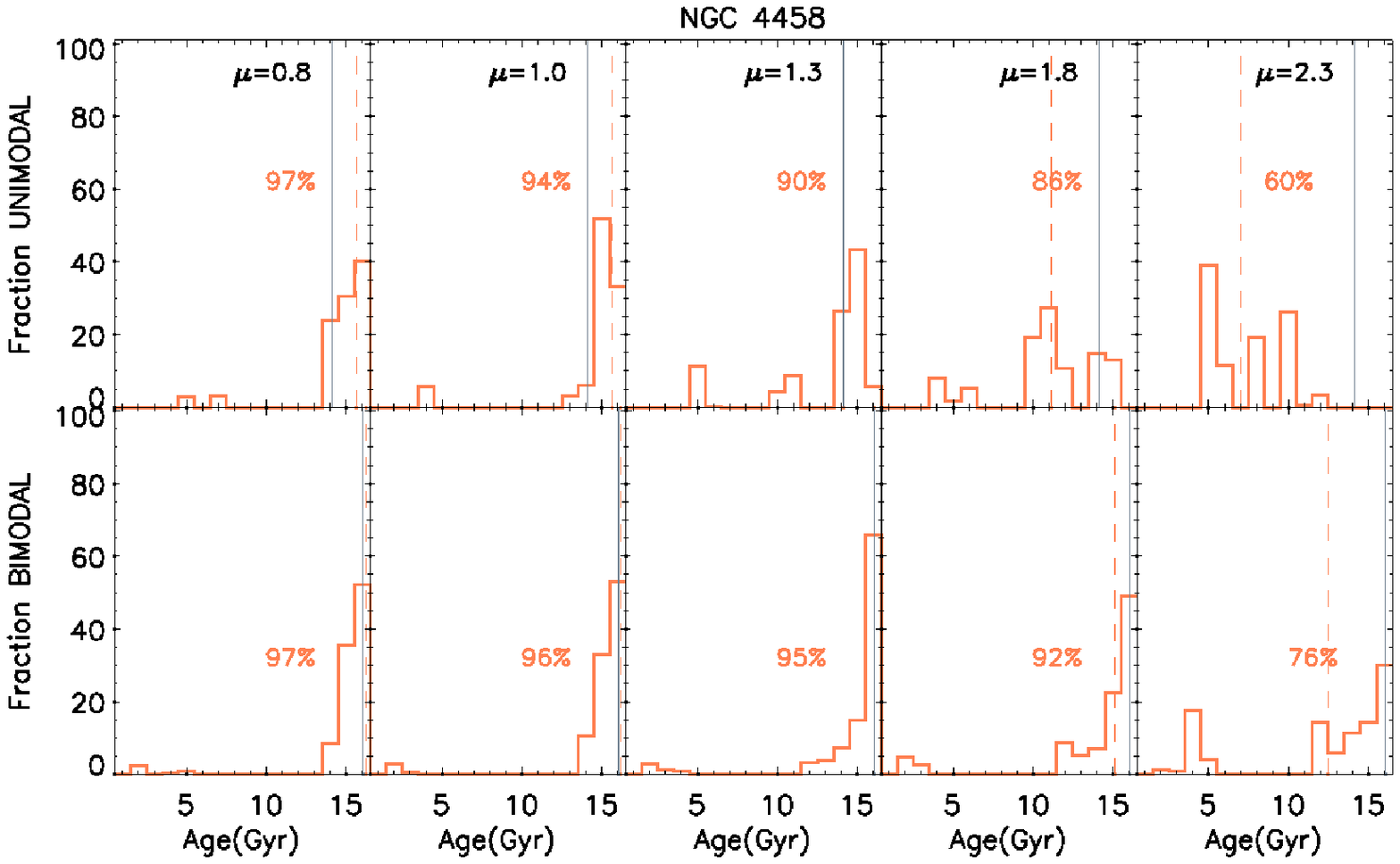}\\
\label{Fig.2}
\caption{Star Formation Histories for the very massive galaxy NGC\,4472, the massive elliptical NGC\,4473, the massive compact galaxy NYU\,685469 and for the low-mass elliptical NGC\,4458. From left to right, the mass-weighted SFHs derived with increasing IMF slope ($\mu$), as indicated on the top of each panel, for the unimodal in the upper row and the bimodal in the lower one. The dashed vertical line corresponds to the mean mass-weighted age derived for the galaxy with each slope, while the solid line represents the mean value derived from the standard $\mu$=\,1.3 slope. The fraction of mass-weighted old populations ($>$5\,Gyr) is quoted in each panel. Note the trend of a decreasing old stellar population contribution when steepening the IMF slope, more remarkably for the massive compacts and the low-mass galaxies.}
\end{figure*}  

\begin{figure*}
\includegraphics[scale=0.65]{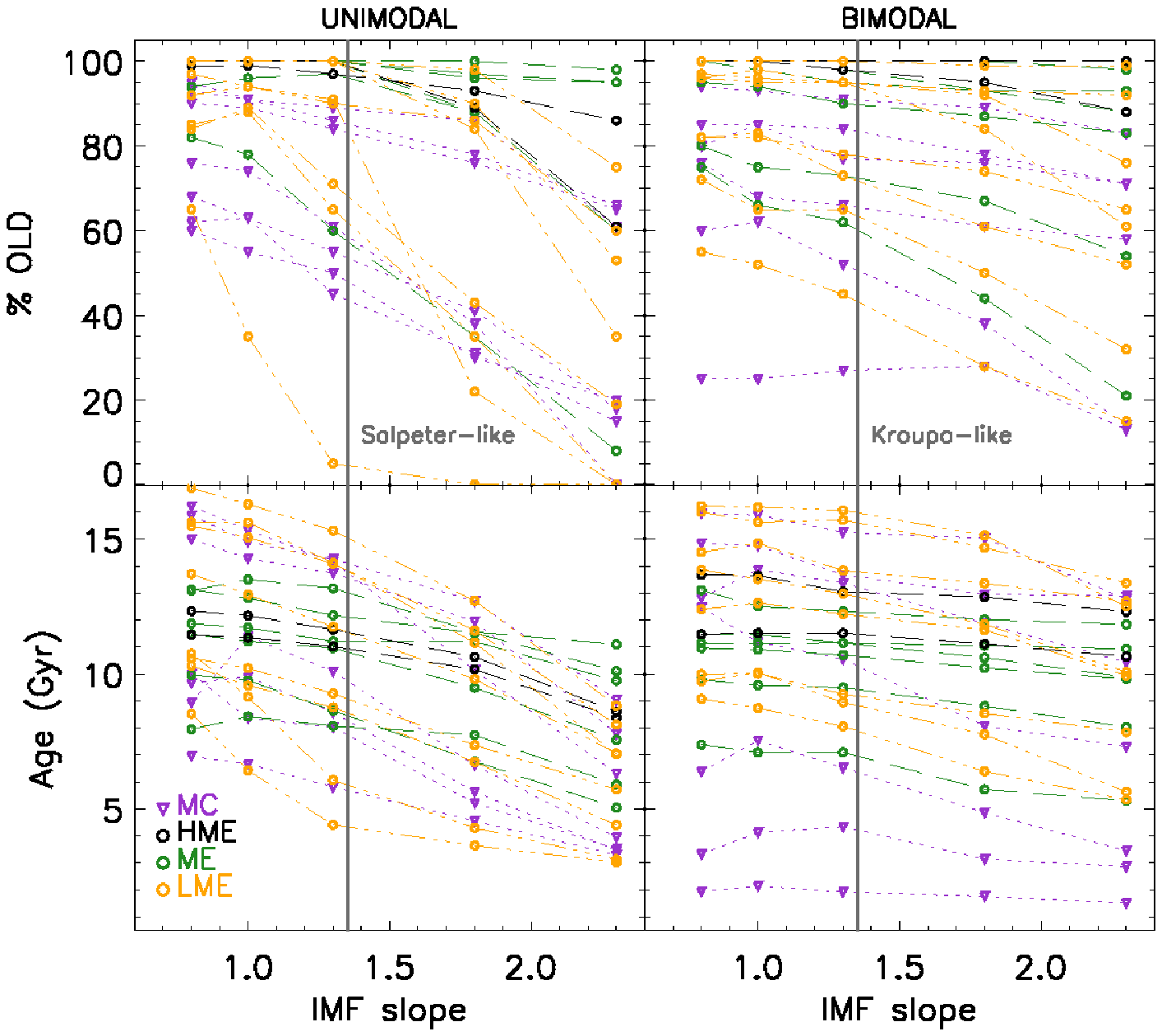}
\label{Fig.3}
\caption{Fraction of old populations ($>$5\,Gyr) and mean ages derived from the mass-weighted SFHs using different IMF slopes. Symbols follow the previous colour code: high massive galaxies in black, massive ellipticals in green, the massive compacts in purple and the low-mass ellipticals in yellow. The vertical solid gray line indicates the position of the standard slope for both IMF shapes ($\mu$=\,1.3). The bimodal case shows milder variations with a decrease of a factor of 1.5, half the variation observed for the unimodal case. }
\end{figure*} 

\subsection{Stellar Masses}  
Stellar masses are computed from the mass-to-light relation corresponding to the SSP mixtures that match best the galaxy spectra, as obtained from the full-spectrum-fitting technique applied here. For our massive compact galaxies, the M/L values were calculated in the r-band of the SDSS. We adopted the {\tt GALFIT} magnitude (\citealt{Peng2002}, \citealt{Peng2010}) as luminosity estimate from AFM12, after correcting for galactic extinction. For the elliptical galaxies we adopted the M/L values in the V band as luminosity estimate. We used the values tabulated for the total magnitude, corrected in the same band, from LEDA (http://leda.univ-lyon1.fr/). Figure 5 shows the dependence of the derived stellar mass with the IMF slope. This is the observable with the largest sensitivity to the assumed IMF shape and slope. It is seen that for the unimodal IMF the stellar mass grows up by a factor of 3 from the standard value $\mu$=\,1.3 to the steepest case $\mu$=\,2.3, with the smallest stellar mass obtained at $\mu\,\sim$\,1.0. For the bimodal IMF shape the impact is significantly milder. It shows a smaller increase of the factor of 1.5, emphasized for the massive galaxies. The trend of increasing mass with a steepening of the IMF slope is analog to the results found in \citet{Cappellari2012a} for the dynamical masses. This finding provides further support for the need of using the appropriate IMF, in order to avoid galaxy misclassifications due to wrong stellar masses. As an example, assume that a sample of massive galaxies will be constructed under the condition: $M_{*}\,\ga 10^{11}\,M_{\sun}$. If the stellar population parameters were derived with a Chabrier IMF, the stellar masses would be underestimated and a fraction of galaxies would be lost.

\begin{figure*}
\includegraphics[scale=0.65]{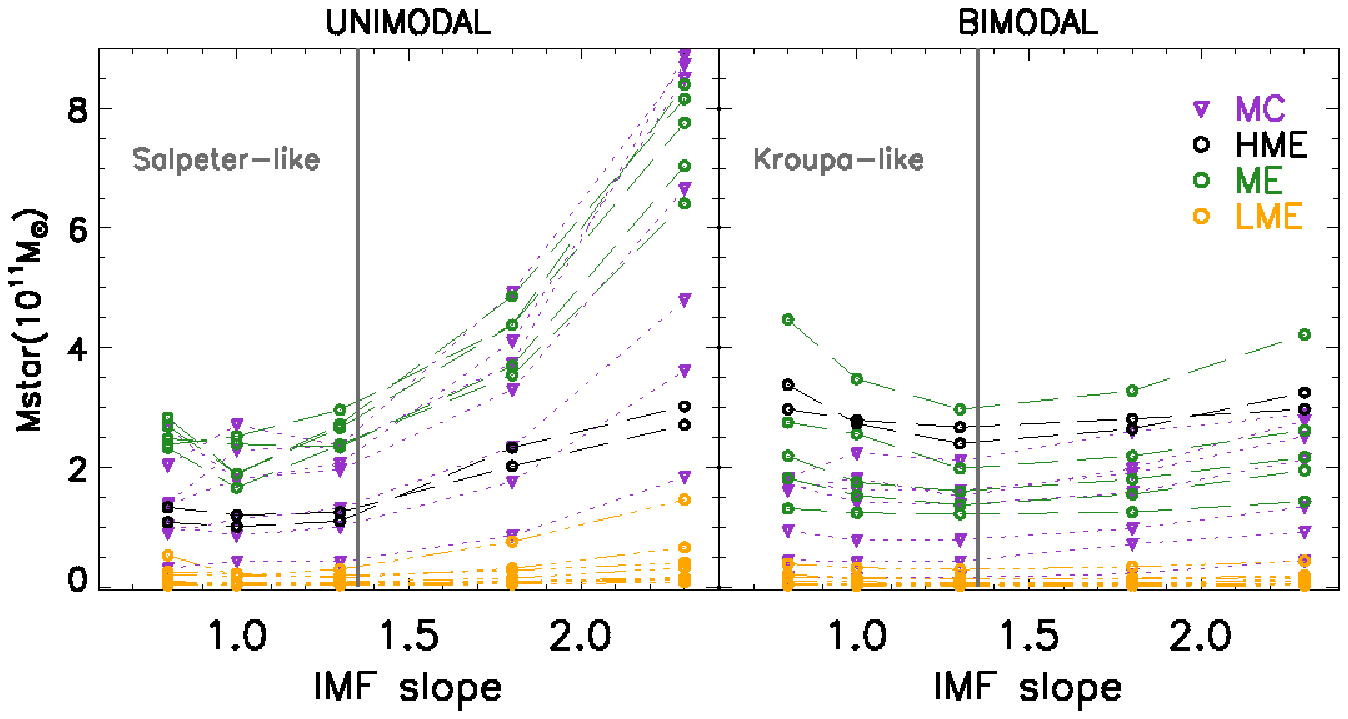}
\label{Fig.4}
\caption{Stellar masses obtained from the derived M/L ratios with different IMF slopes. Note the large difference between unimodal and bimodal shapes, in particular for the steepest slopes. The stellar masses increase by a factor of 4 in the unimodal case, while it only accounts by a factor of 1.5 in the bimodal case. }
\end{figure*}

\begin{table*}
\centering
\caption{Stellar properties with $\mu(\sigma)$ }           
\label{table:2}     
\begin{tabular}{c| c c c c c c c c}   
\hline\hline      
   & \multicolumn{2}{|c}{$\mu(\sigma)$} & \multicolumn{2}{|c}{Age (Gyr)} & \multicolumn{2}{|c}{$\%$old} & \multicolumn{2}{|c}{M$_{star}$ ($\times$\,10$^{11}M_{\sun}$)}\\    
\hline 
 GALAXY ID  & un & bi & un & bi & un & bi & un & bi\\
\hline 
 NYU\,54829  & 0.8 (0.8) & 0.8 (0.8) & 6.9  & 6.3  & 68  & 60  & 0.91 & 0.94 \\      
 NYU\,321479 & 1.3 (1.5) & 2.3 (2.2) & 13.7 & 10.4 & 84  & 71  & 2.35 & 2.52 \\     
 NYU\,685469 & 1.3 (1.4) & 1.8 (1.9) & 8.5  & 1.9  & 50  & 0   & 1.01 & 0.72 \\    
 NYU\,796740 & 1.3 (1.4) & 1.8 (1.9) & 15.3 & 15.0 & 89  & 89  & 2.36 & 2.60 \\     
 NYU\,890167 & 1.8 (1.7) & 2.3 (2.3) & 6.6  & 2.8  & 41  & 13  & 0.86 & 0.45 \\      
 NYU\,896687 & 1.8 (1.6) & 2.3 (2.3) & 12.6 & 12.9 & 78  & 71  & 4.11 & 2.78 \\       
 NYU\,2434587& 1.3 (1.4) & 1.8 (1.9) & 8.0  & 8.1  & 61  & 61  & 1.96 & 1.58 \\
 NGC\,4365  & 1.8 (1.9) & 2.8 (2.7) & 10.6 & 12.3 & 89  & 100 & 2.02 & 3.25 \\
 NGC\,4472  & 2.3 (2.2) & 3.0 (3.1) & 8.4  & 10.6 & 86  & 88  & 3.02 & 2.97 \\
 NGC\,2329  & 1.8 (1.6) & 2.3 (2.2) & 7.7  & 5.9  & 88  & 21  & 4.38 & 1.95 \\ 
 NGC\,4473  & 1.3 (1.2) & 1.8 (1.7) & 10.9 & 10.6 & 100 & 93  & 2.38 & 3.28 \\
 NGC\,4621  & 1.8 (1.6) & 2.3 (2.3) & 11.2 & 11.8 & 96  & 98  & 4.38 & 1.43 \\ 
 NGC\,4697  & 1.0 (1.1) & 1.3 (1.5) & 9.7  & 9.4  & 78  & 74  & 1.94 & 2.65 \\ 
 NGC\,5812  & 1.3 (1.5) & 1.8 (2.0) & 12.8 & 11.1 & 100 & 95  & 1.89 & 1.98 \\  
 NGC\,4239  & 0.8 (0.8) & 0.8 (0.8) & 10.7 & 12.4 & 92  & 82  & 0.07 & 0.05 \\  
 NGC\,4339  & 0.8 (0.8) & 0.8 (0.8) & 10.6 & 9.9  & 84  & 82  & 0.53 & 0.38 \\
 NGC\,4387  & 0.8 (0.8) & 0.8 (0.8) & 13.7 & 13.8 & 100 & 96  & 0.03 & 0.03 \\
 NGC\,4458  & 0.8 (0.8) & 0.8 (0.8) & 15.6 & 16.2 & 97  & 97  & 0.07 & 0.06 \\ 
 NGC\,4464  & 0.8 (0.8) & 0.8 (0.8) & 16.3 & 16.0 & 100 & 100 & 0.19 & 0.20 \\
 NGC\,4467  & 0.8 (0.8) & 0.8 (0.8) & 15.5 & 14.5 & 100 & 96  & 0.06 & 0.04 \\ 
 NGC\,4489  & 0.8 (0.8) & 0.8 (0.8) & 8.5  & 9.7  & 65  & 72  & 0.09 & 0.09 \\ 
 NGC\,4551  & 0.8 (0.8) & 0.8 (0.8) & 10.3 & 9.1  & 85  & 55  & 0.25 & 0.17 \\  
\hline                                  
\end{tabular}\\
{Stellar populations parameters estimated from the derived SFHs using the prescribed IMF slope, according the the galaxy velocity dispersion ($\mu(\sigma)$): (1) galaxy ID; (2-3) The selected $\mu(\sigma)$ for each galaxy, both for the unimodal and the bimodal IMF. We consider the closest slope in our model (0.8, 1.0, 1.3, 1.8 or 2.3) compared to the value obtained following F13 equation (quoted within the parenthesis). Note that all the low-mass galaxies have $\mu(\sigma)$=\,0.8, because F13 equation is valid for galaxies with $\sigma\,>$150 $\rm{km\,s^{-1}}$; (4-5) Mean mass-weighted ages; (6-7) Fraction of old stellar populations ($>$5\,Gyr); (8-9) Stellar mass derived from the SSP mixture resulting from the derived SFH.}
\end{table*}

\section{NGC\,4489: baby galaxy or non-standard IMF?}                                                                                                                   
We have seen in Section 3.2 that by tuning the IMF slope related to the velocity dispersion of the galaxy, according to recent evidence in this direction, we tend to derive more similar SFHs among galaxies of different masses. However such a choice does not change drastically the results one would infer from adopting a standard-universal IMF (e.g. Salpeter or Kroupa). We find that the most massive galaxies are nevertheless old, despite IMF-modulated young population contributions. However, here we will show that there are galaxies for which the use of a standard IMF would lead to surprising results.\\
Here we focus on the low-mass elliptical NGC\,4489. This galaxy has been previously reported to have young mean luminosity-weighted ages from line-strength studies ($\sim$3-5\,Gyr; e.g: \citealt{Gorgas1997}, \citealt{Terlevich2002}, \citealt{Caldwell2003}, \citealt{Tantalo2004}, Y06).
However, this does not necessarily imply that the galaxy is composed of genuinely young stellar populations, as the old component can be masked by the luminosity of a tiny fraction of young stars on top of it. The mass-weighted SFH reveals the composition of NGC\,4489, as presented in Figure 5. It shows, like in Fig.2, the derived SFHs with different IMF slopes. From the Salpeter panel (unimodal $\mu$=\,1.3) one would infer that NGC\,4489 is a genuinely young object with almost no contribution from stellar populations older than 5\,Gyr (less than 5$\%$), i.e. a ``\textit{baby elliptical galaxy}''. However, considering a slightly flatter IMF slope as claimed for an elliptical galaxy of
such small velocity dispersion (e.g. \citealt{Cappellari2012a}, F13), an old component appears, which contributes $\sim\,$65$\%$ to the total mass. This result is in much better agreement with those obtained in Section 3.2.\\ 
Do other galaxies like NGC\,4489 exist in the Local Universe that can be considered as \textit{baby elliptical galaxies} assuming a standard IMF slope? Out of the SDSS-based LB10 sample, we selected those galaxies with a mean mass-weighted age younger than $\leq$\,5\,Gyr (Salpeter IMF). Only 11 \textit{baby elliptical galaxies} were found, with a similar behavior to NGC\,4489, as shown in Fig. 6. Once again, the apparent extreme youth of these galaxies results from the inadequate use of a standard-universal IMF.\\

\begin{figure}
\includegraphics[scale=0.5]{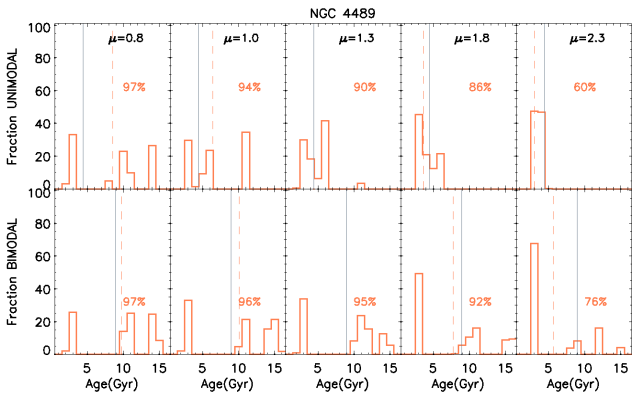}
\label{Fig.5}
\caption{Derived SFHs for NGC\,4489 for increasing IMF slopes, as quoted within each panel as in Figure 2. Note that for a standard Salpeter IMF (third panel) this object is genuinely young, while a slight flattening of the IMF slope gives a SFH with a large contribution of old stellar populations.}
\end{figure} 

\begin{figure}
\centering
\includegraphics[scale=0.6]{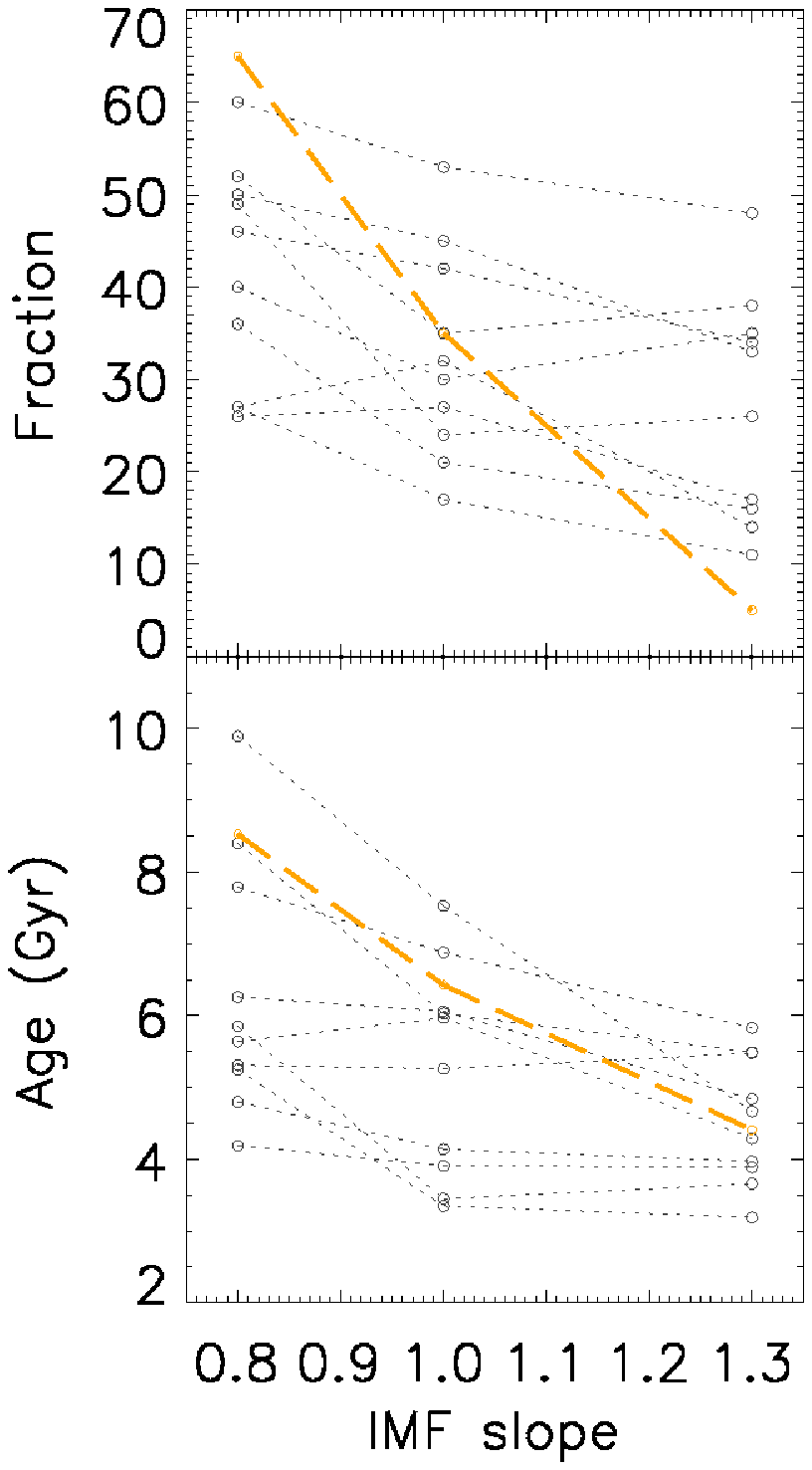}
\label{Fig.6}
\caption{Fraction of old stellar populations ($>$\,5\,Gyr) and mean age (in Gyr) estimated from the derived SFH using the unimodal IMF of varying slope for a subsample of 11 \textit{baby elliptical galaxies} from LB10. The Virgo galaxy NGC\,4489, with a high-quality spectrum, is highlighted in yellow. As for NGC\,4489, these \textit{baby elliptical galaxies} always show an almost genuinely young stellar population when a Salpeter IMF is adopted.}
\end{figure}

\section{Conclusions}
We have gone one steep further on the discussion about the universality of the IMF. Although in this work we do not attempt to constrain the IMF, we explore and quantify the impact of varying its slope and shape on various galaxy properties (M/L, SFHs, M$_{star}$ and  mean ages). These are summarized as follows: 
\begin{itemize}
 \item The choice of the IMF shape (unimodal \textit{vs} bimodal) results in different stellar population parameter estimates, in particular for the steepest slopes, as seen from the M/L ratios. The unimodal (bimodal) IMF shape shows stronger (milder) deviations on the derived properties with respect to the standard IMF.
 \item  We find a minimum on the M/L for an IMF slope that depends slightly on the age of the SSP: $\mu$\,=\,0.8 for young SSPs, increasing towards $\mu$\,=\,1.0 for the oldest stellar populations in the unimodal case. For the bimodal this minimum is located at $\mu$\,=\,1.0. This implies that at $\mu\sim$\,0.8-1.0, the stellar mass derived from the SFHs will be minimized. In the unimodal case the use of steeper IMF slopes increases the derived stellar masses by a factor of 4 (from $\mu$=\,0.8 to $\mu$=\,2.3), while the increase is by a factor of 1.5 in the bimodal one. This is the most sensitive parameter to the IMF variations.
 \item IMF slopes steeper than Salpeter tend to produce SFHs with smaller contributions of old populations. This effect decreases the mean age of the galaxy, both in mass and light. This effect is observed for both IMF shapes analyzed here, although in the unimodal case the decreases are more drastic, by a factor of up to 3 (from $\mu$=\,0.8 to $\mu$=\,2.3). It shows milder versions, up to 1.5, for the bimodal case. 
 \item The choice of an IMF slope according to the velocity dispersion of the galaxy seems to provide more constrained results. While standard IMFs produce rather mass (and age) dependent SFH types, the use of the $\mu$-$\sigma$ relation produces comparable SFHs for the ETG family. The most massive ellipticals, for which we adopt slopes steeper than Salpeter, present a small, but not negligible contribution of relatively younger stellar populations. Within this scenario,
massive galaxies would no longer be completely dead and passively evolving old objects, supporting recent claims in this direction. Adopting flatter IMF slopes for the low-mass galaxies will render more similar, but still different, SFHs to the massive ones. However, the SFHs derived for our compact massive galaxies, which show unprecedented large contributions of young stellar populations, do not change significantly if the IMF slope is varied. We also find galaxies with apparently no contribution from old stellar populations using a Salpeter IMF (\textit{baby elliptical galaxies}). However, the use of flatter slopes, implied by their low velocity dispersion, uncover a conspicuous old population. 
 \end{itemize}

These results point out that a variation of the IMF slope might solve, or at least alleviate, some of the problems encountered when analyzing galaxies of varied masses. Contrary to earlier warnings against the use of non standard IMFs, the present study provides new caveats on the assumption of a universal IMF, such as the questionable existence of early-type galaxies with virtually no  contributions of old stellar populations, or the misclassified non-massive galaxies due to mass underestimation. Moreover, the complex SFHs found for most massive galaxies when using steeper IMFs open new means to explain their formation and evolution. \\

\section*{Acknowledgments}

We thank the referee for the detailed comments and relevant suggestions that have certainly contributed to improve the final version. We also would like to thank P. S\'anchez-Bl\'azquez and J. Falc\'on-Barroso for very useful discussions. This work has been supported by the Programa Nacional de Astronom{\'{\i}}a y Astrof{\'{\i}}sica of the Spanish Ministry of Science and
Innovation under grant AYA2010-21322-C03-02. Based on observations made with: WHT/ING Telescope in El Roque the los Muchachos Observatory of the Instituto de Astrof{\'{\i}}sica de Canarias, La Palma.

\bibliography{IMF_SFH_AFM.v3}
\bibliographystyle{mn2e}

\appendix

\section{Testing our method for deriving the SFHs with {\tt STARLIGHT}}
Full-spectrum fitting codes are known to be slightly sensitive to the initial conditions, such as the spectral range (e.g. blue \textit{vs} red), or masked features (e.g. potential emission lines). We have tested the robustness of our results by varying these initial conditions, while keeping the IMF shape and slope fixed to a unimodal $\mu$=1.3.\\
Altogether, these tests reveal that the trends and estimates derived from the relatively short spectral range of the long-slit galaxy spectra are robust.

\subsection{Masking test}
Three masking configurations were used, involving different features on the spectra of four of our galaxies, as illustrated in the upper panel of figure A1: (a) orange -- no mask is used; (b) blue -- we mask all the features in the spectra, both potential emission and Balmer-absorption lines; (c) green -- we only mask the Balmer-absorption lines. The resulting mean luminosity-weighted
ages could be used to provide a more quantitative estimate of the uncertainties derived from the method. We find that these mean ages, which are indicated for each panel, vary insignificantly, by less than 2$\%$.

\begin{figure*}
\includegraphics[scale=0.8]{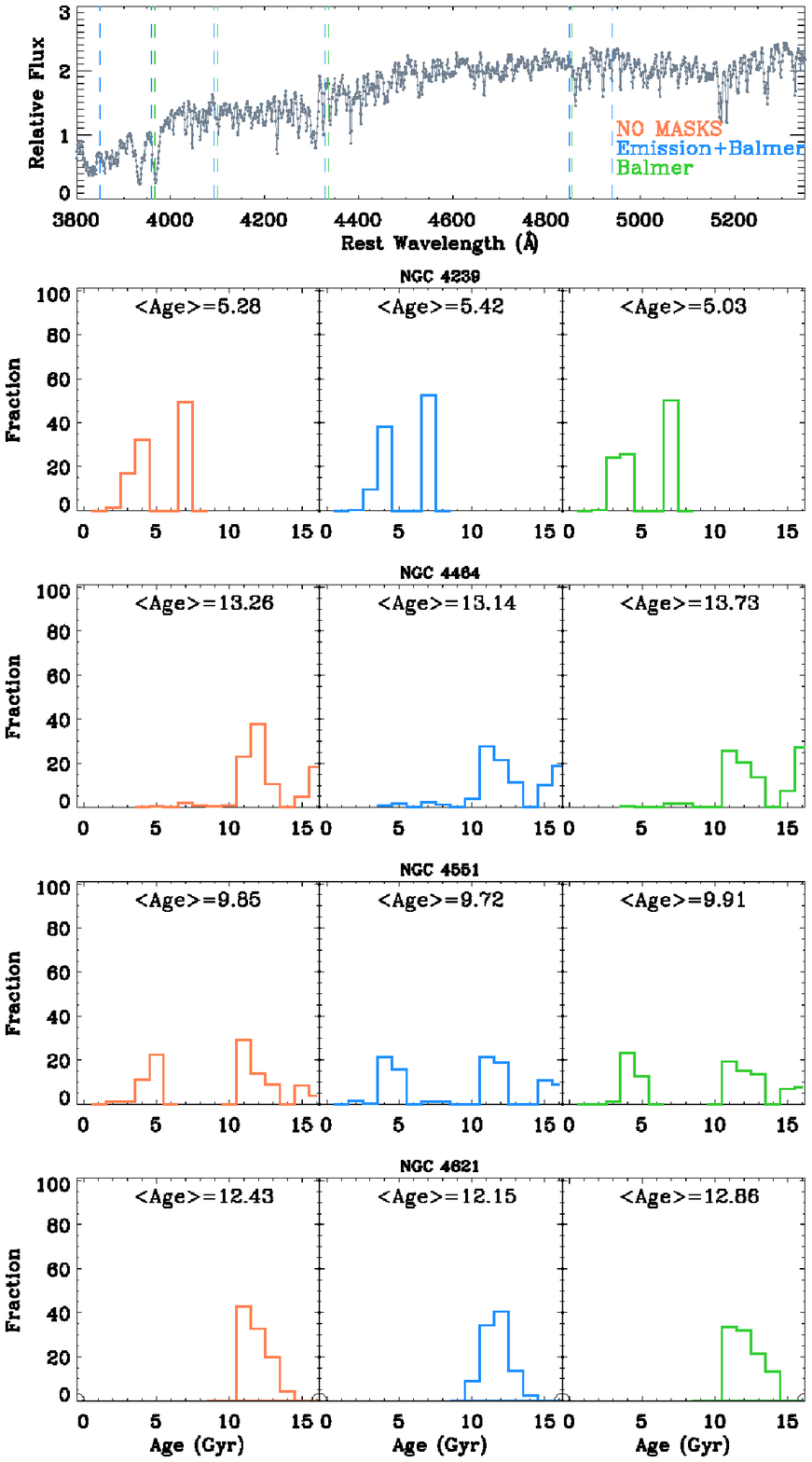}\\  
\label{Fig.A1}
\caption{Different masking configurations employed for the spectrum-fitting are indicated on the upper panel. The vertical blue and green dashed lines mark the masked lines for each configuration. The SFHs derived for four elliptical galaxies are shown in the following rows of panels. No masking (left), masking emission and Balmer lines (center) and only masking the Balmer lines (right).
The mean mass-weighted age is shown for each SFH, giving us a quantitative estimate of the uncertainties of this method.} 
\end{figure*} 

\subsection{Short \textit{vs} Long spectral range}
We have compared the mean ages derived from the relatively short range (3800-5300$\,\rm\AA{}$) of the high quality long-slit spectra of our sample of galaxies with those derived from the full SDSS spectral range. The latter covers bluer and redder wavelengths (3500 to 8500$\,\rm\AA{}$ ), better sampling the different contributions from the various stellar populations. For this test we
have selected a set of ETGs from the sample of LB10, covering a wide range in ages. Moreover, as the compact massive galaxies were initially selected from the DR6 SDSS database \citep{Trujillo2009}, we also compared their results. The luminosity-weighted estimates show a rather good agreement, with deviations on the derived mean age smaller than 8$\%$, whereas the mass-weighted estimates show deviations on the mean age of around 12$\%$. The larger variations
are found for the galaxies with the oldest ages, as expected for the errors associated to this age regime.

\begin{figure*}
\centering
\includegraphics[scale=0.85]{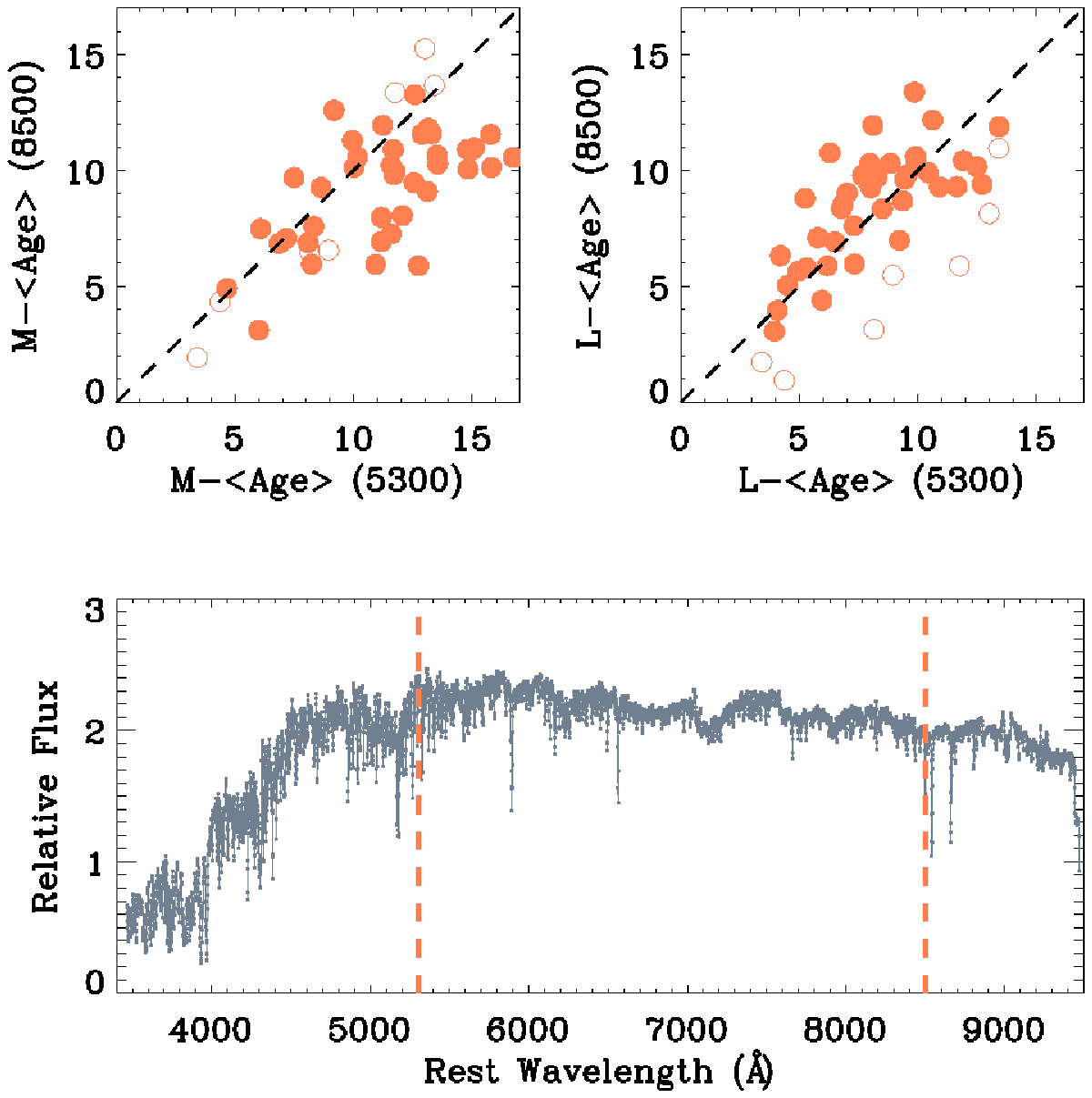}
\label{Fig.A2}
\caption{Mean mass- and light-weighted ages (left and right, respectively) obtained from the SFHs derived using the short and large spectral ranges. Filled symbols correspond to the galaxies from the LB10 sample, while open symbols correspond to the compact massive galaxies with SDSS data. The vertical lines on the spectra indicate these two spectral ranges. }
\end{figure*}

\subsection{Broadband photometry \textit{vs} Spectrum fitting}
Colour variations with the IMF slope have been already reported (e.g. \citealt{Ricciardelli2012}, \citealt{Pforr2012}). However, the effect of the colours in our short spectral range is almost negligible. We compared the SDSS photometric colours with those obtained from the SFHs derived from the short spectral range of these galaxies in LB10. To obtain the synthetic colours we use
the user-friendly web-tool facility dedicated for this purpose in the MILES web-page ``\textit{Get spectra from a SFH}''. This facility provides us with the indices, masses and colours for our derived SFHs. Figure A3 shows that the agreement is remarkable between the derived and observed \textit{g-r} and \textit{r-i} colours. The web-tool does not provide us with the magnitude in the
\textit{u} band, because the MIUSCAT SEDs do not cover the whole spectral range required to measure this filter (see \citealt{Ricciardelli2012}).\\

\begin{figure}
\centering
\includegraphics[scale=0.7]{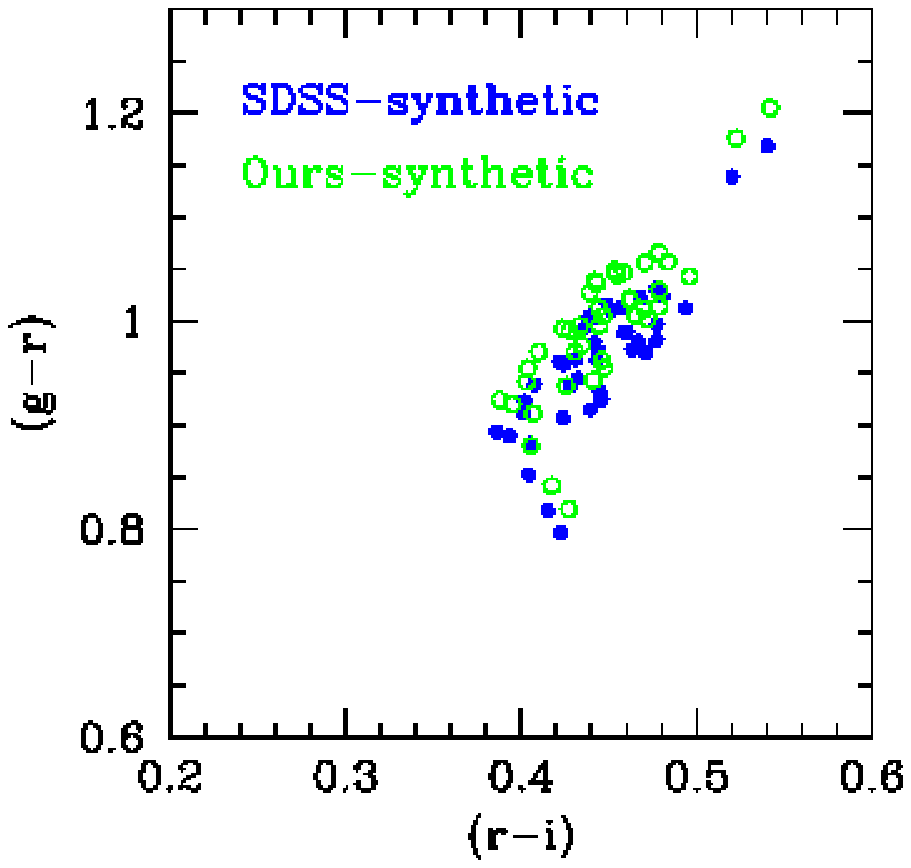}
\label{Fig.A3}
\caption{\textit{g-r} and \textit{r-i} colours derived from our SFHs on the short spectral range (open green circles) compared to those reported by the SDSS (blue filled circles).}
\end{figure}

\section{SFHs derived for the sample of high-quality spectra}
In this Section we show the SFHs derived for all the galaxies of our sample with high-quality spectra. We do not include here the galaxies already shown in Fig.2 and 5. The first six figures correspond to the SFHs of the massive compact galaxies. The next six show the SFHs of the massive and very massive ellipticals and the last six correspond to the low-mass ellipticals. As in Fig. 2, the panels show, from left to right, the derived SFHs obtained when using models with increasing IMF slope. In the top panels we show the results when adopting a unimodal IMF, whereas in the bottom panels we show the bimodal case. The dashed vertical line corresponds to the mean mass-weighted age derived for the galaxy with each slope, while the solid line represents the mean value derived from the standard $\mu$=\,1.3 slope.

\begin{figure*}
        \includegraphics[scale=0.50]{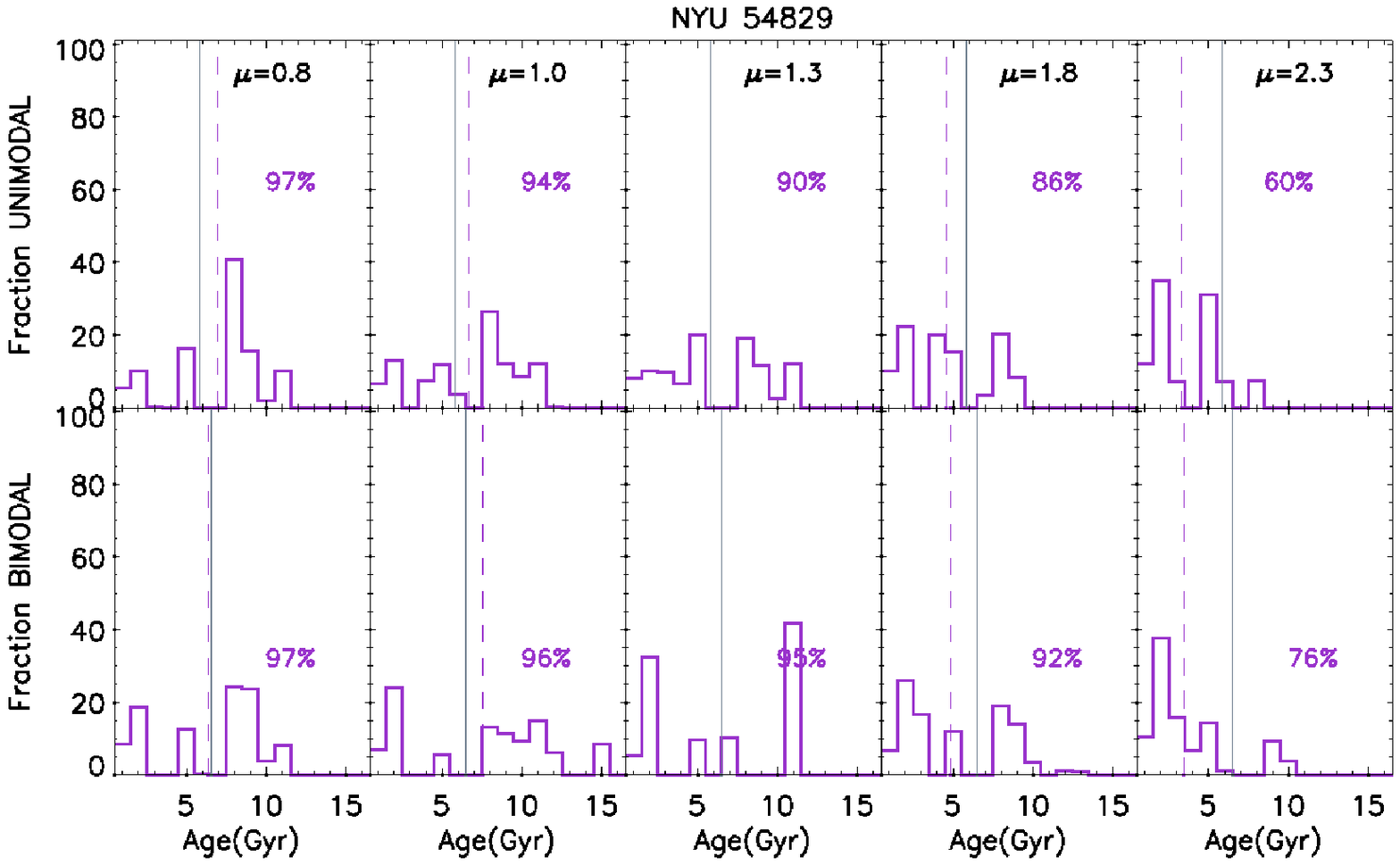}  
        \includegraphics[scale=0.50]{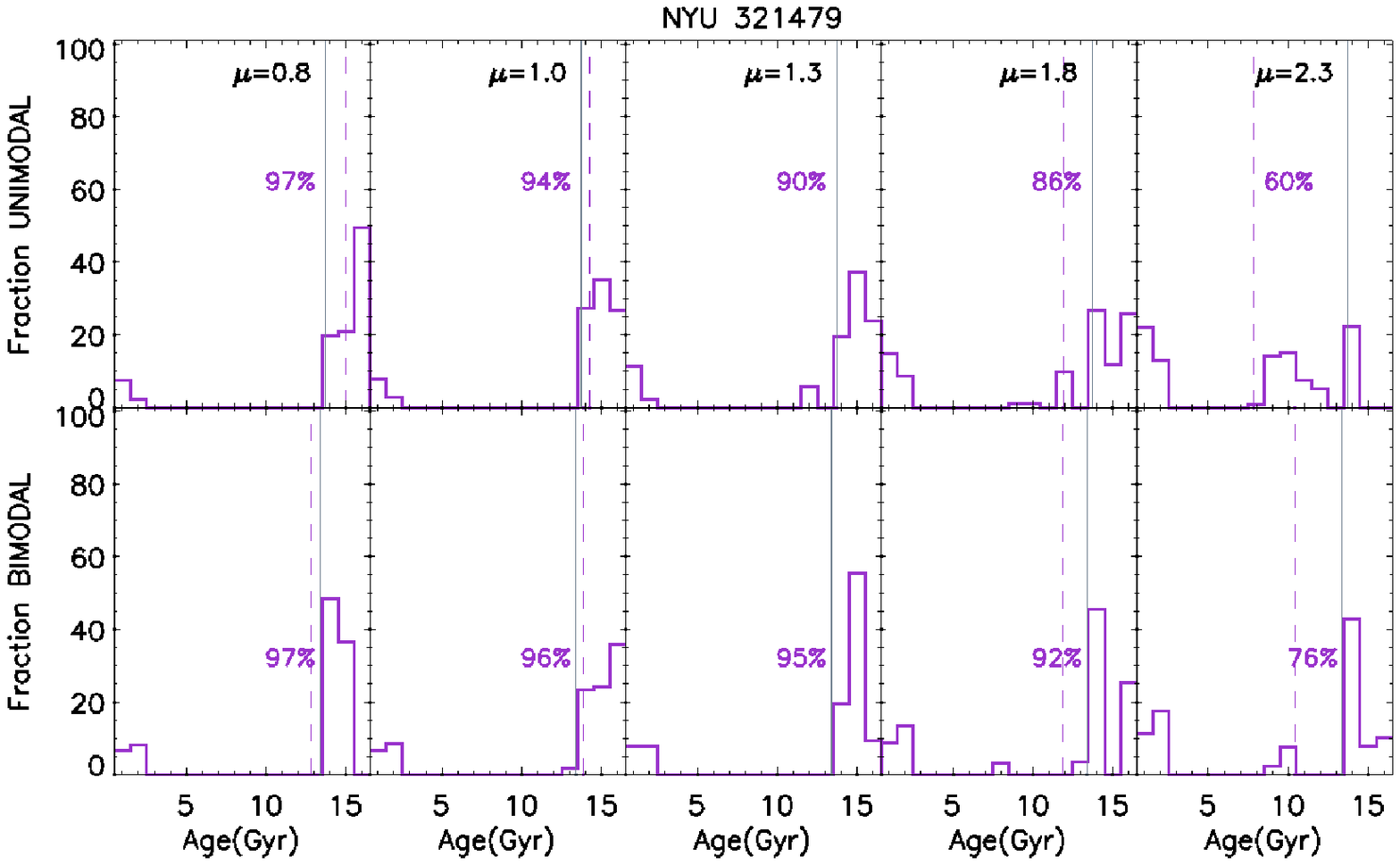}\\ 
        \includegraphics[scale=0.50]{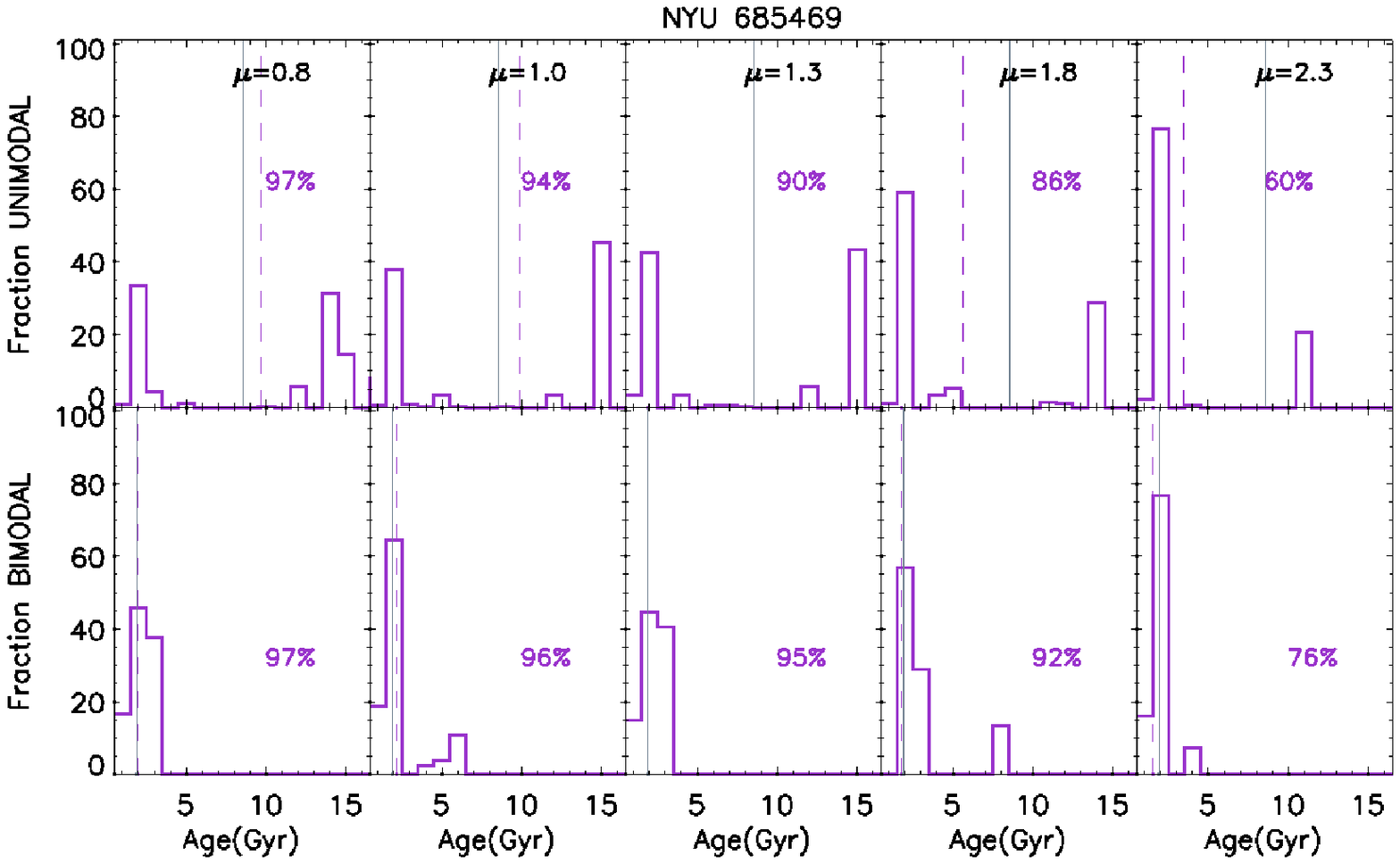} 
        \includegraphics[scale=0.50]{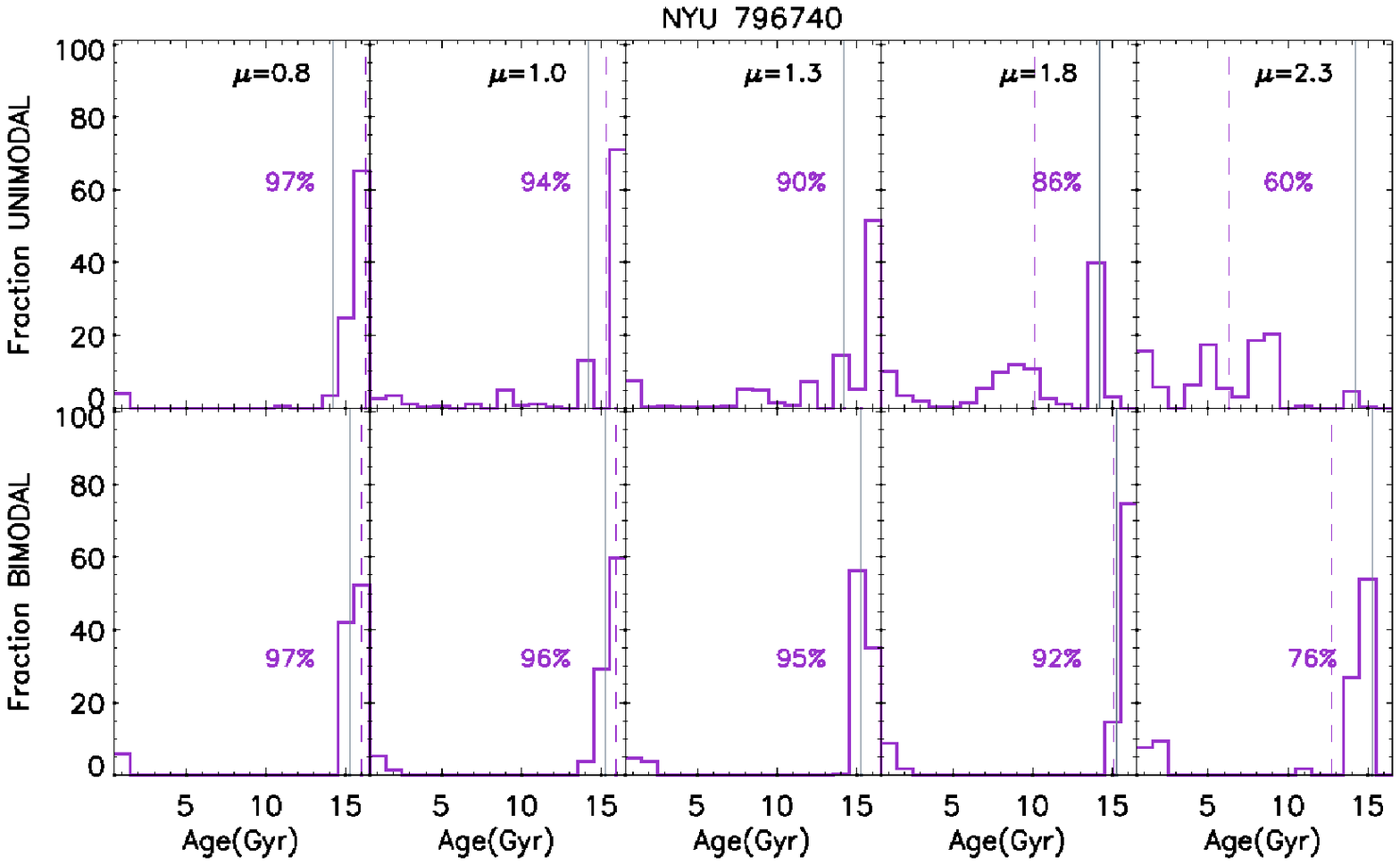}\\ 
        \includegraphics[scale=0.50]{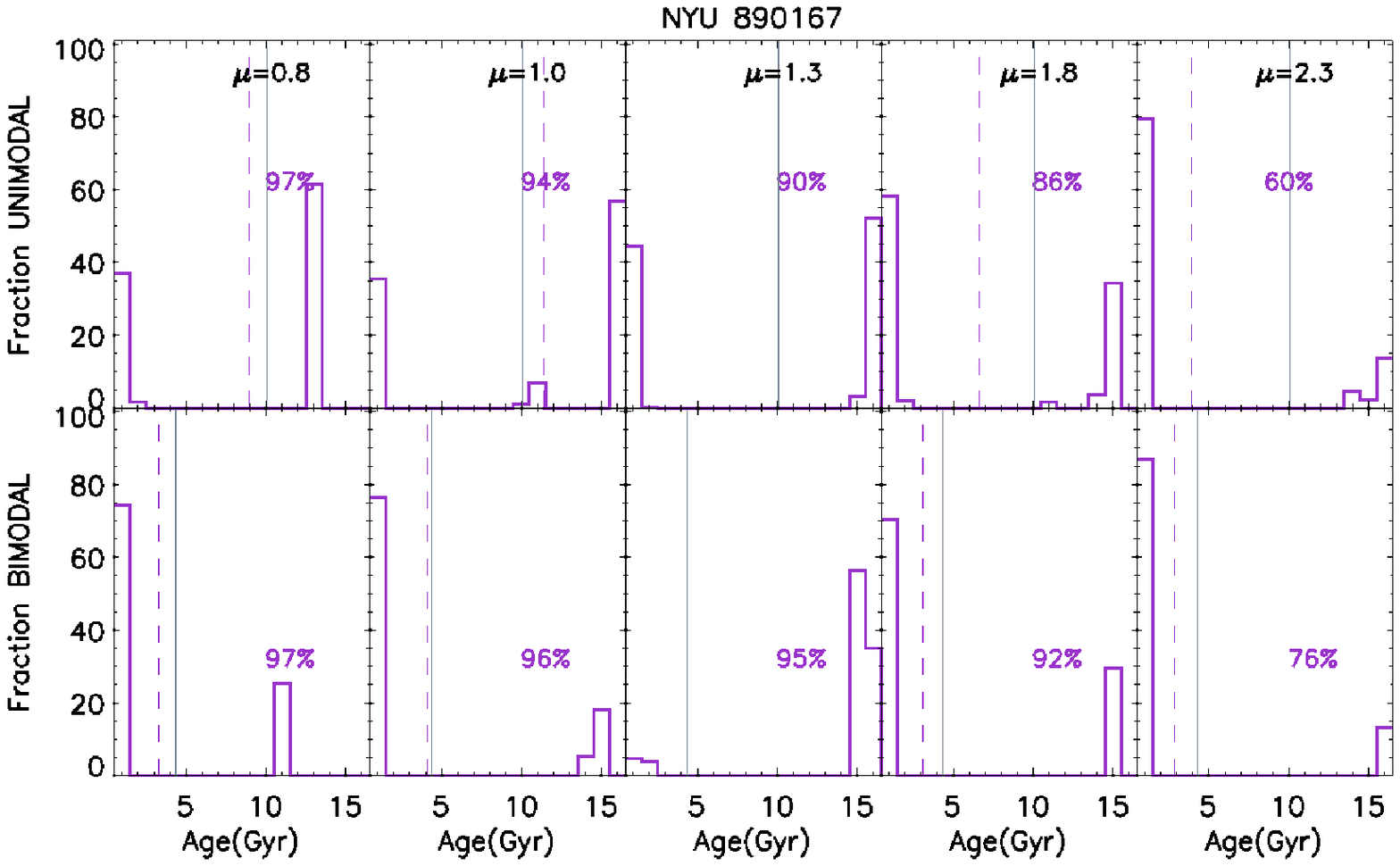}
        \includegraphics[scale=0.50]{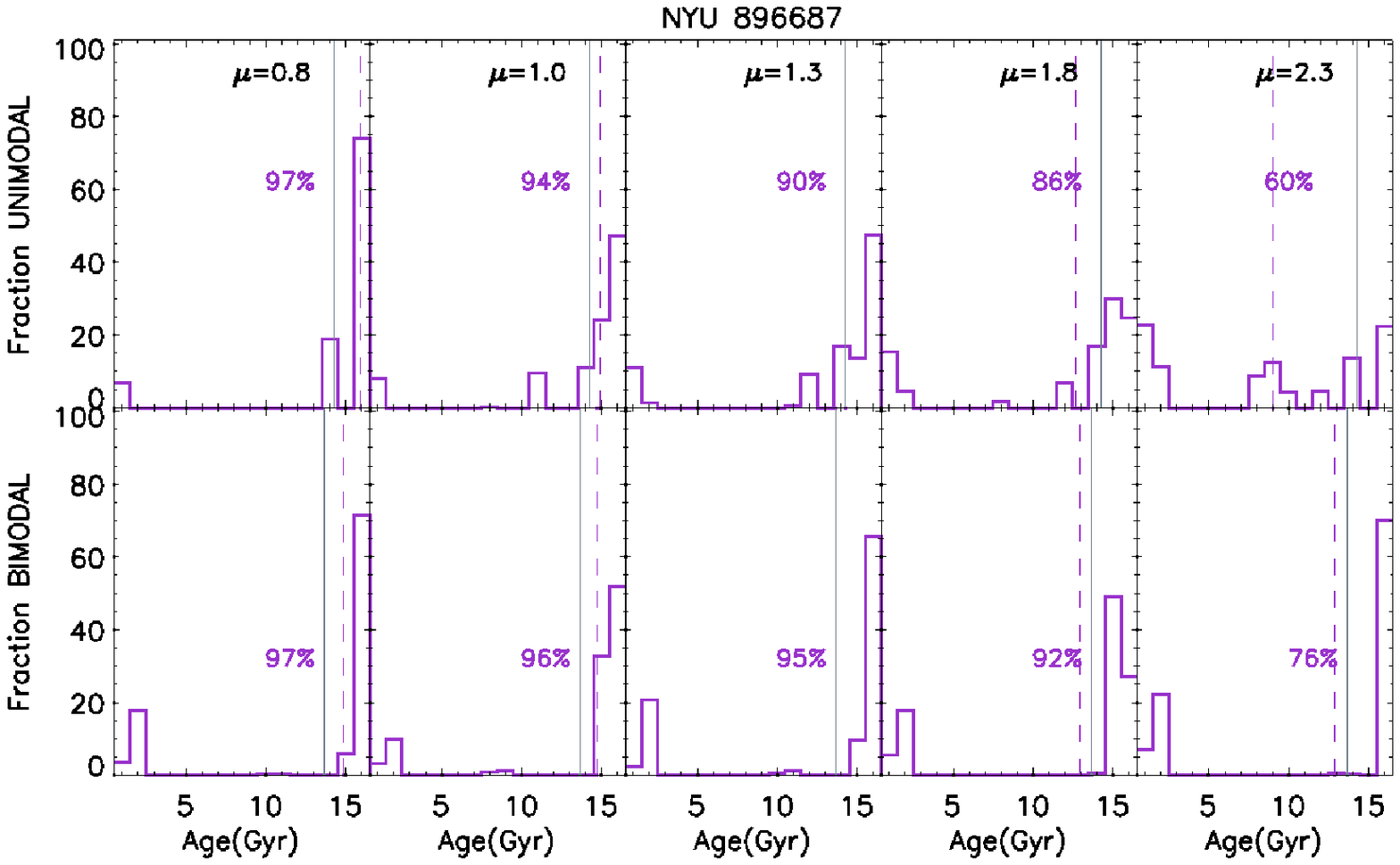}\\
\label{Fig.B1}
\caption{Star formation histories derived for the galaxies in our high-quality sample. The IMF slope of the adopted base models increase from left to right as quoted on each plot, for the unimodal in the upper row and for the bimodal in the lower one. The dashed vertical line indicates the derived mean mass-weighted age with each slope, while the solid line represents the mean value obtained from the standard IMF slope $\mu$=\,1.3}. As for the previous plots, the SFH is colour-coded by galaxy type: black for the high mass ellipticals, green for the massive ellipticals, purple for the compact massive galaxies and yellow for the low-mass ellipticals. A trend where the contribution of the old stellar populations decrease with increasing IMF
slope is clearly seen for all the galaxies. 
\end{figure*} 

\begin{figure*}
        \includegraphics[scale=0.50]{ngc4472_histo.eps}
        \includegraphics[scale=0.50]{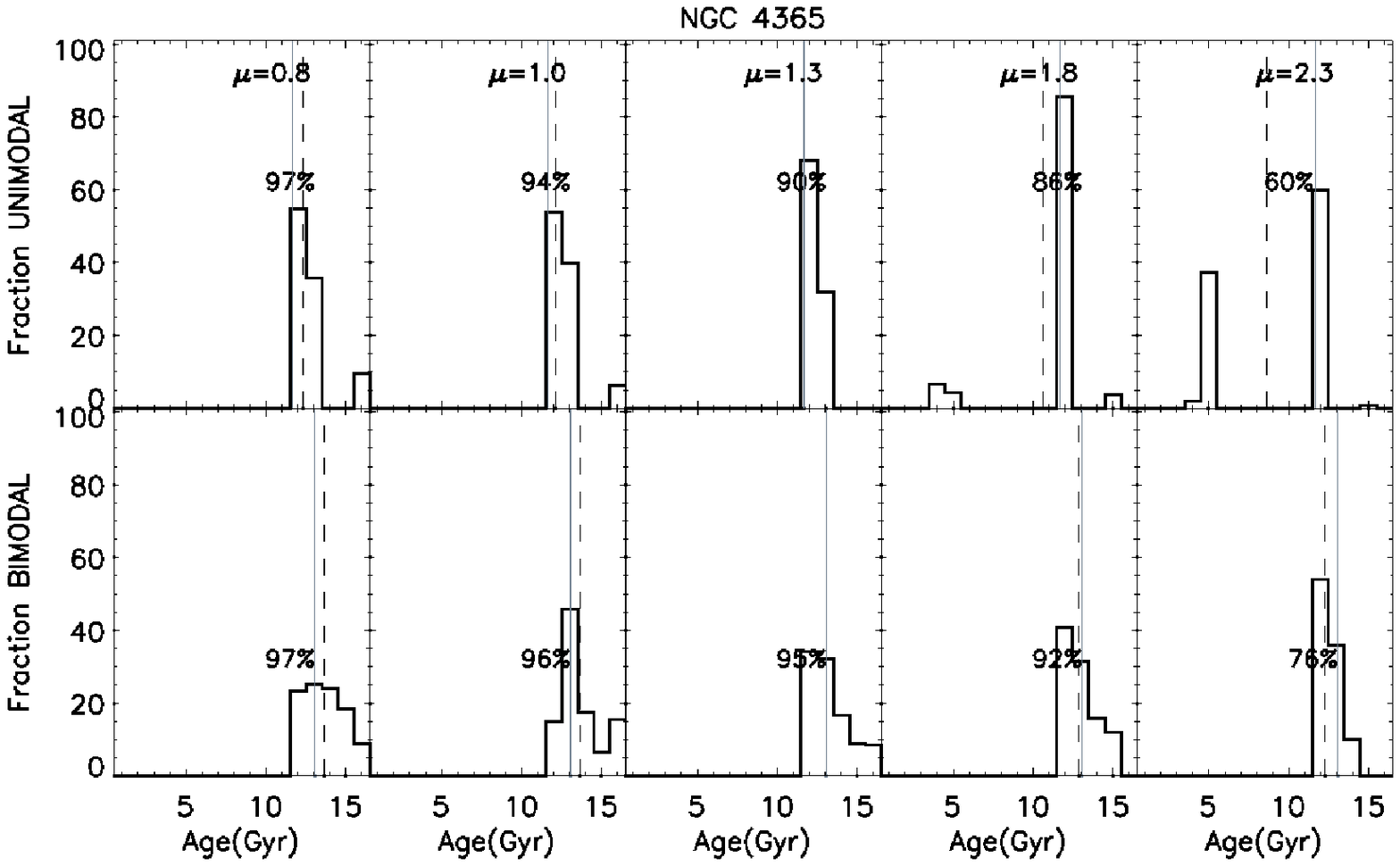}\\ 
        \includegraphics[scale=0.50]{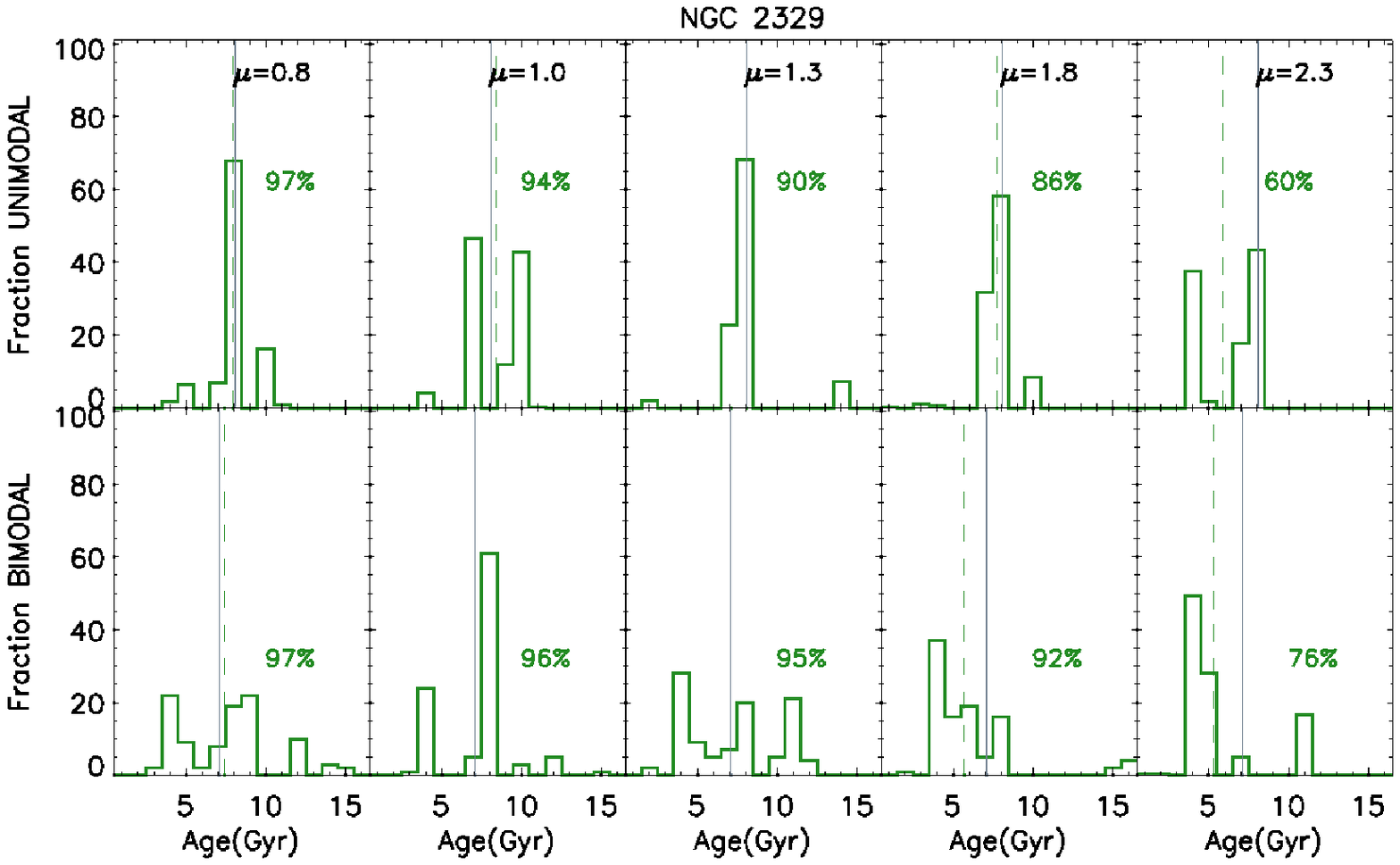} 
        \includegraphics[scale=0.50]{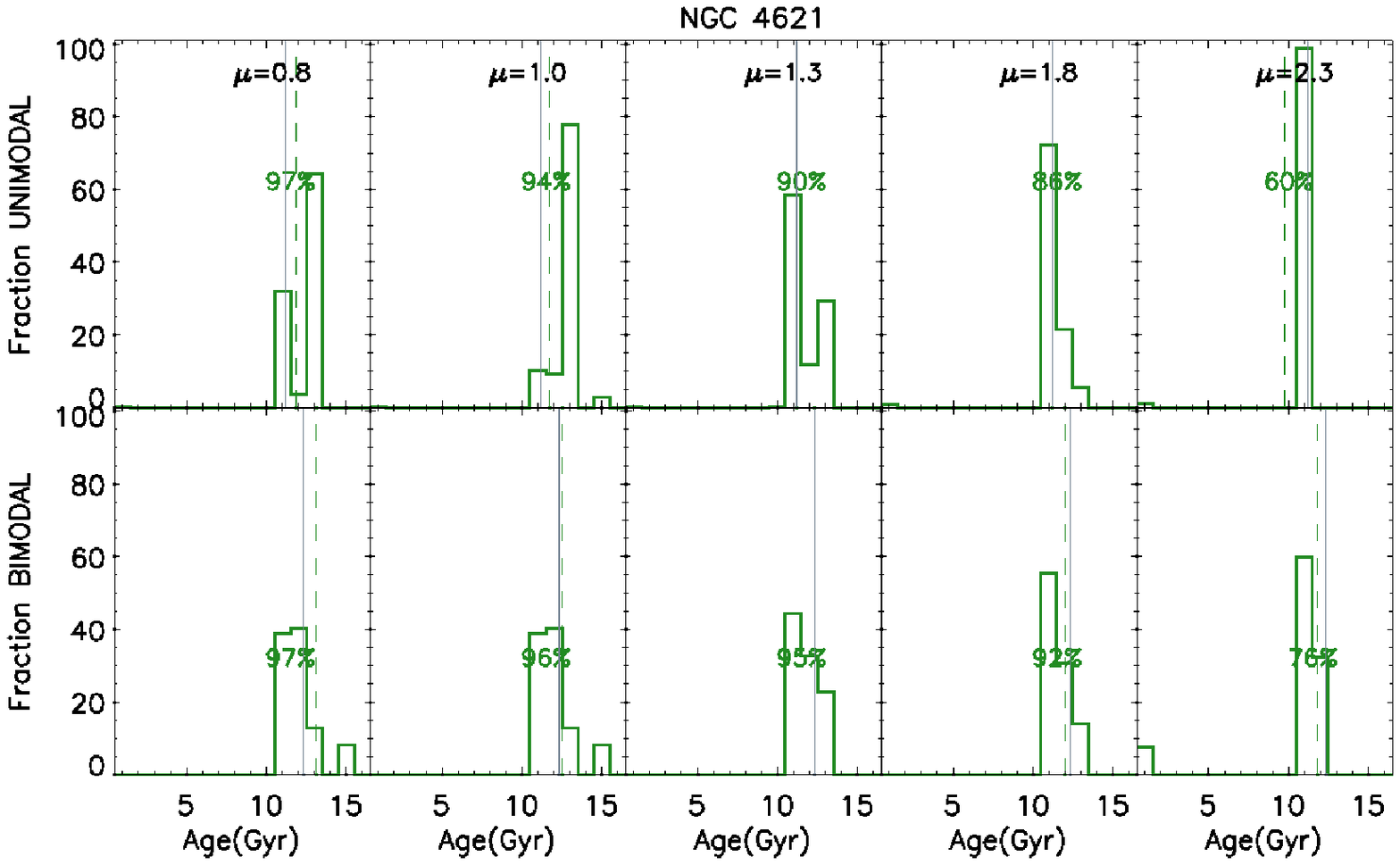}\\ 
        \includegraphics[scale=0.50]{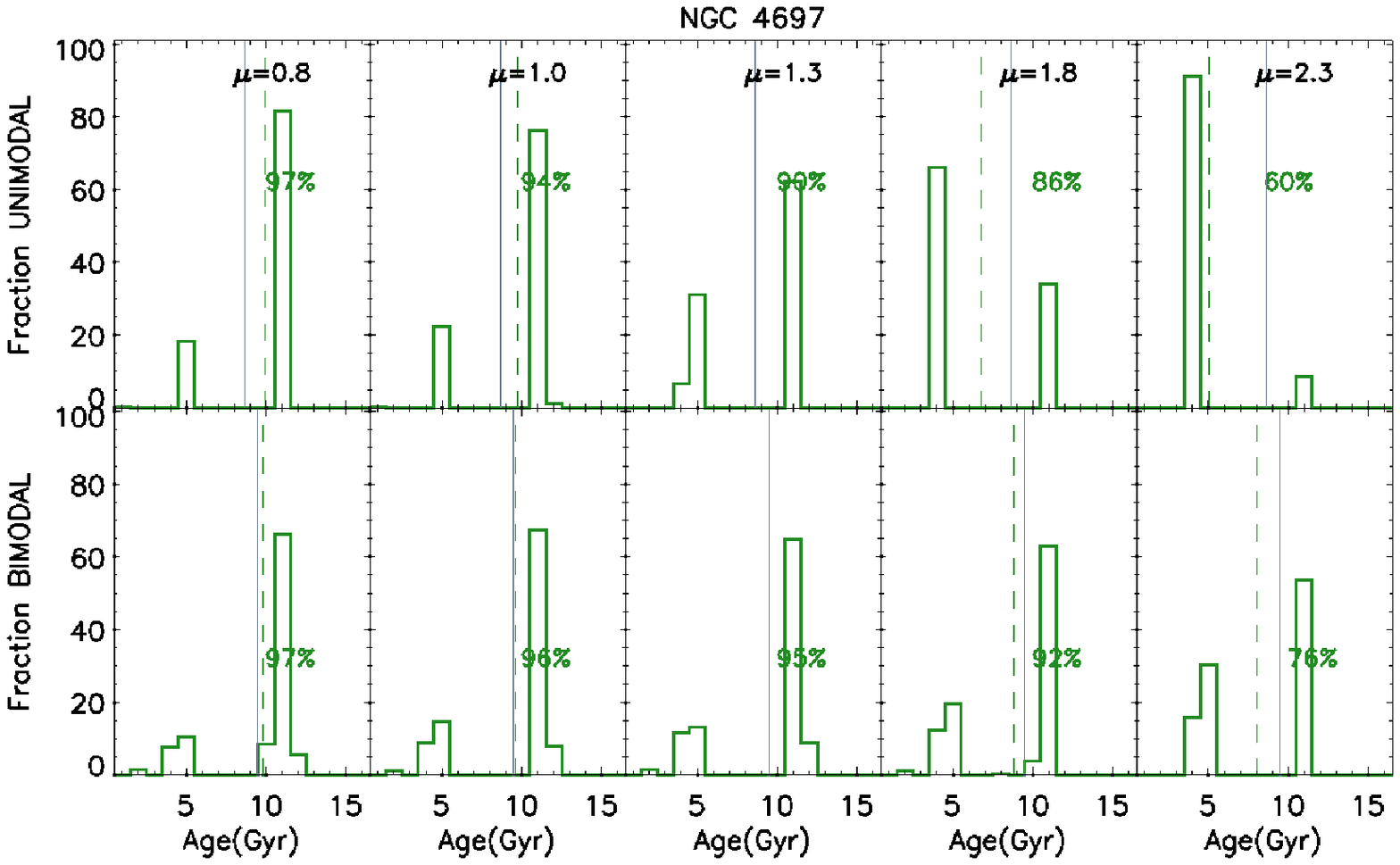}  
        \includegraphics[scale=0.50]{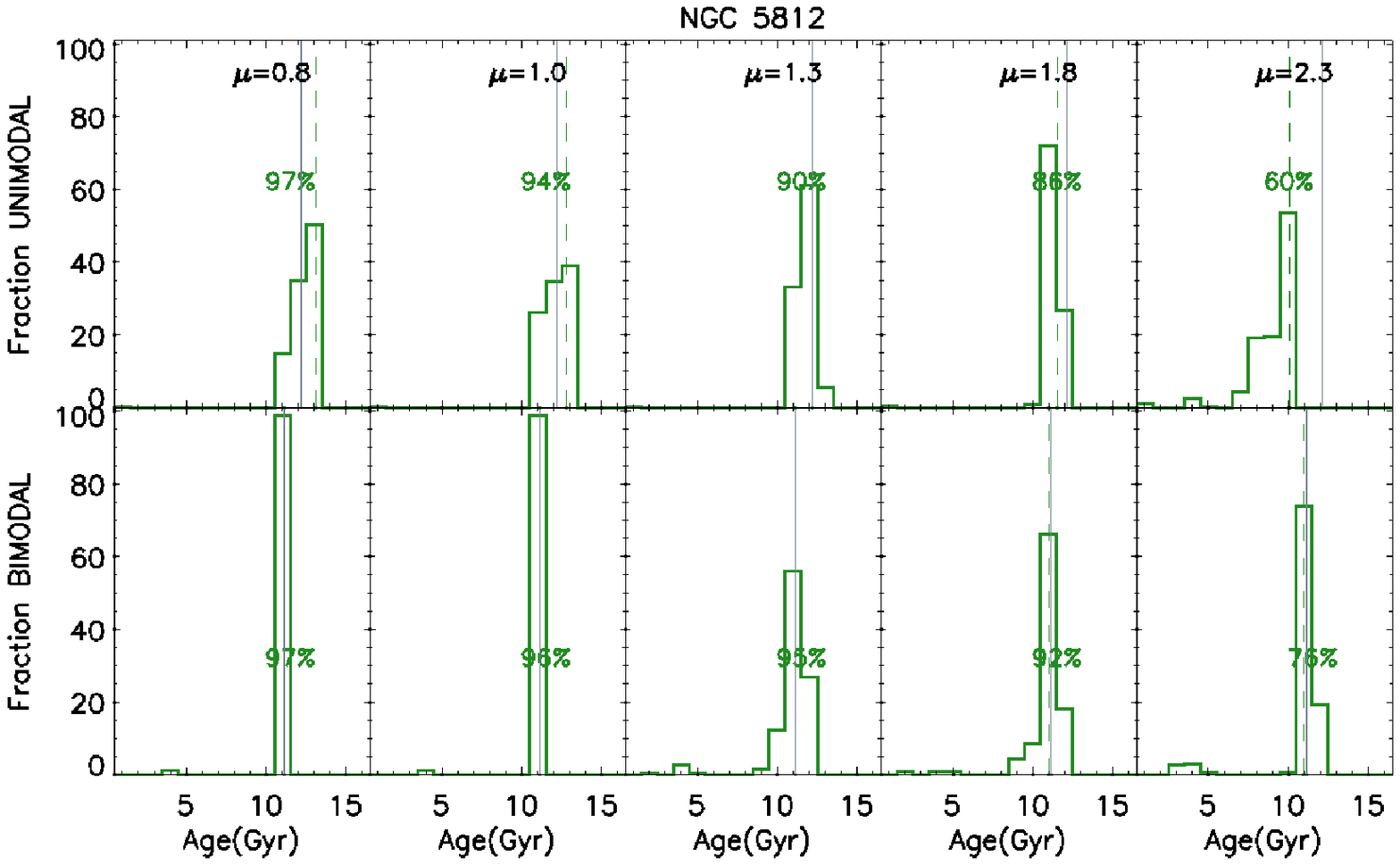}\\
\caption{Fig. B1 continued} 
\end{figure*} 

\begin{figure*}
        \includegraphics[scale=0.50]{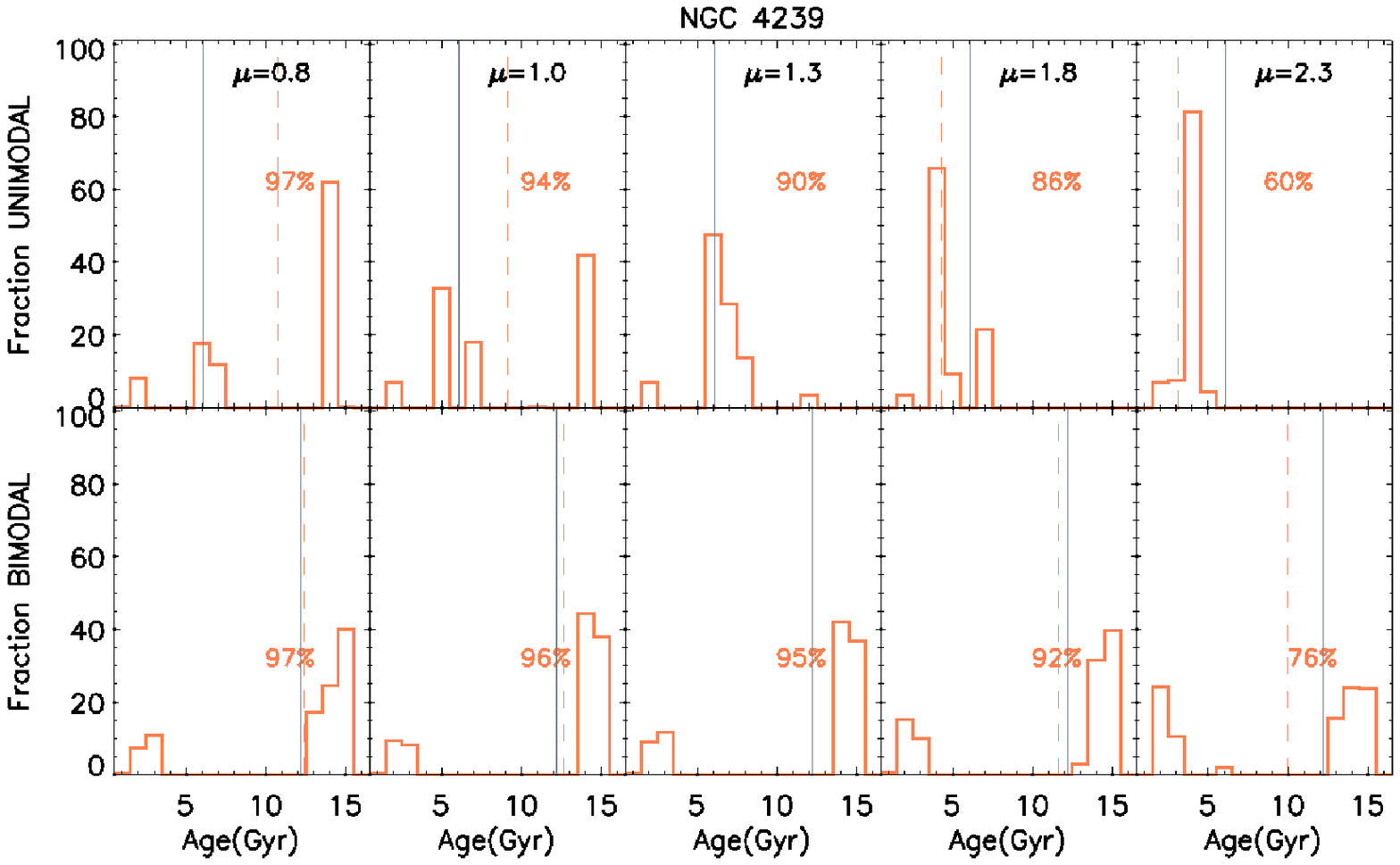} 
        \includegraphics[scale=0.50]{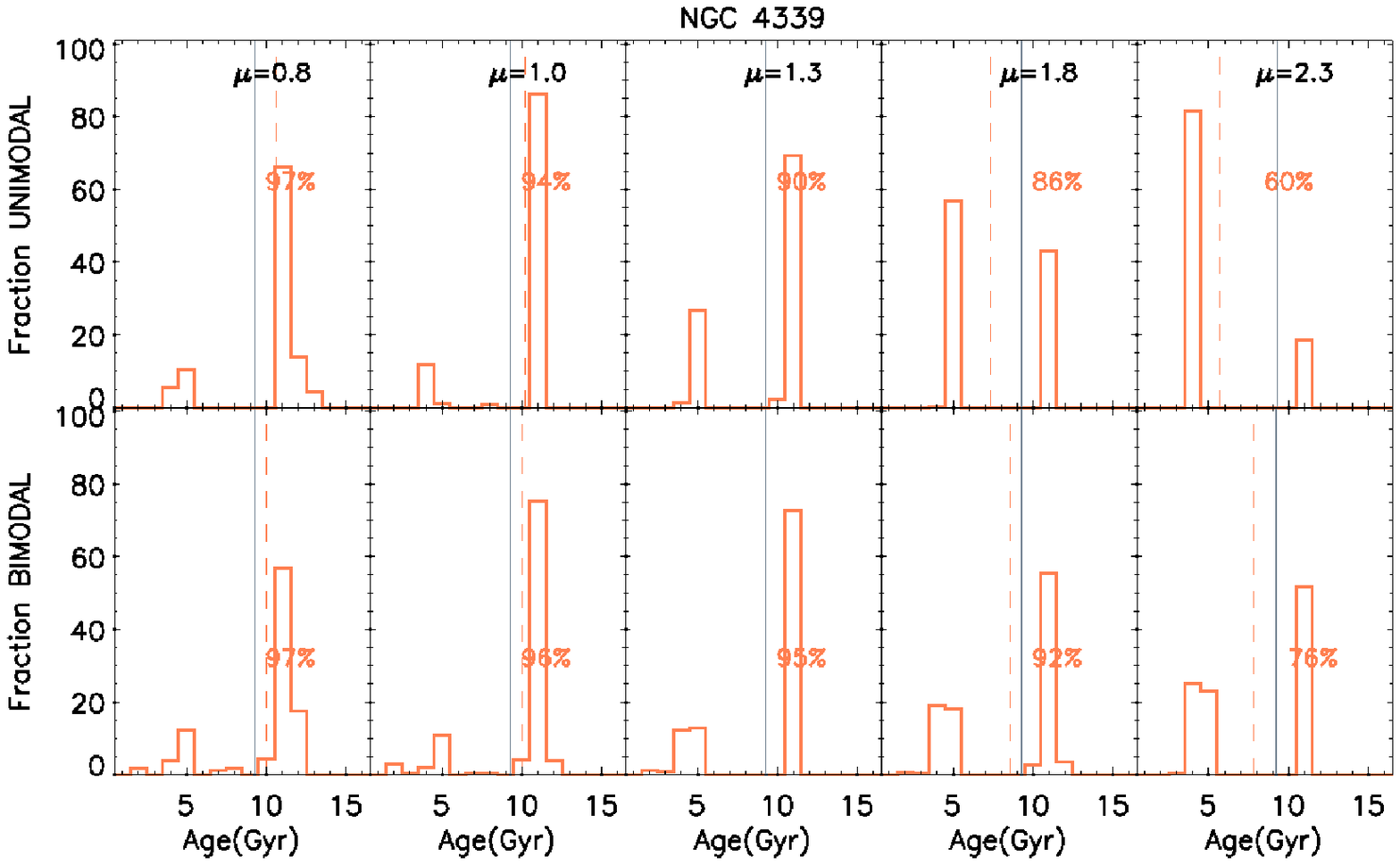}\\
        \includegraphics[scale=0.50]{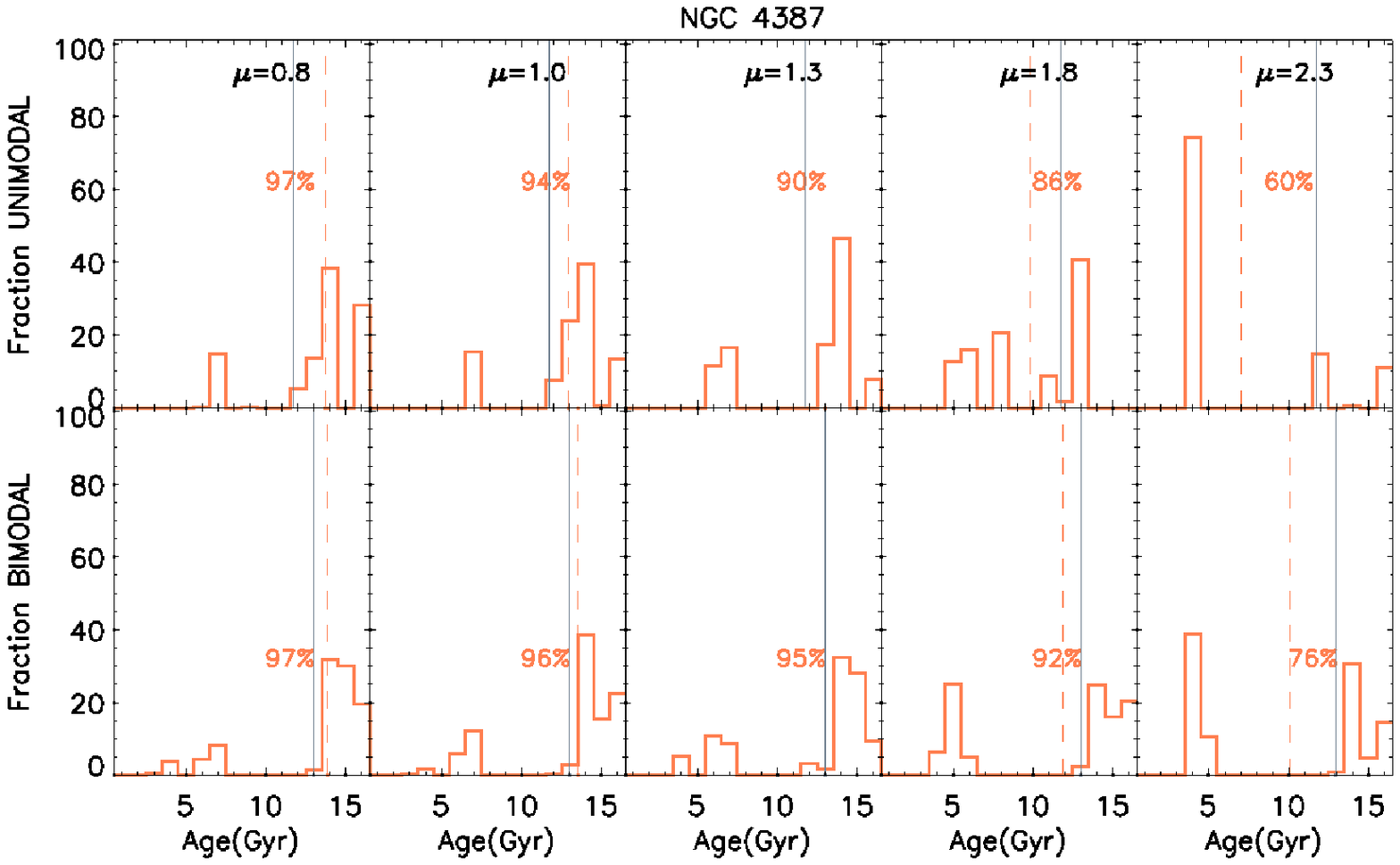} 
        \includegraphics[scale=0.50]{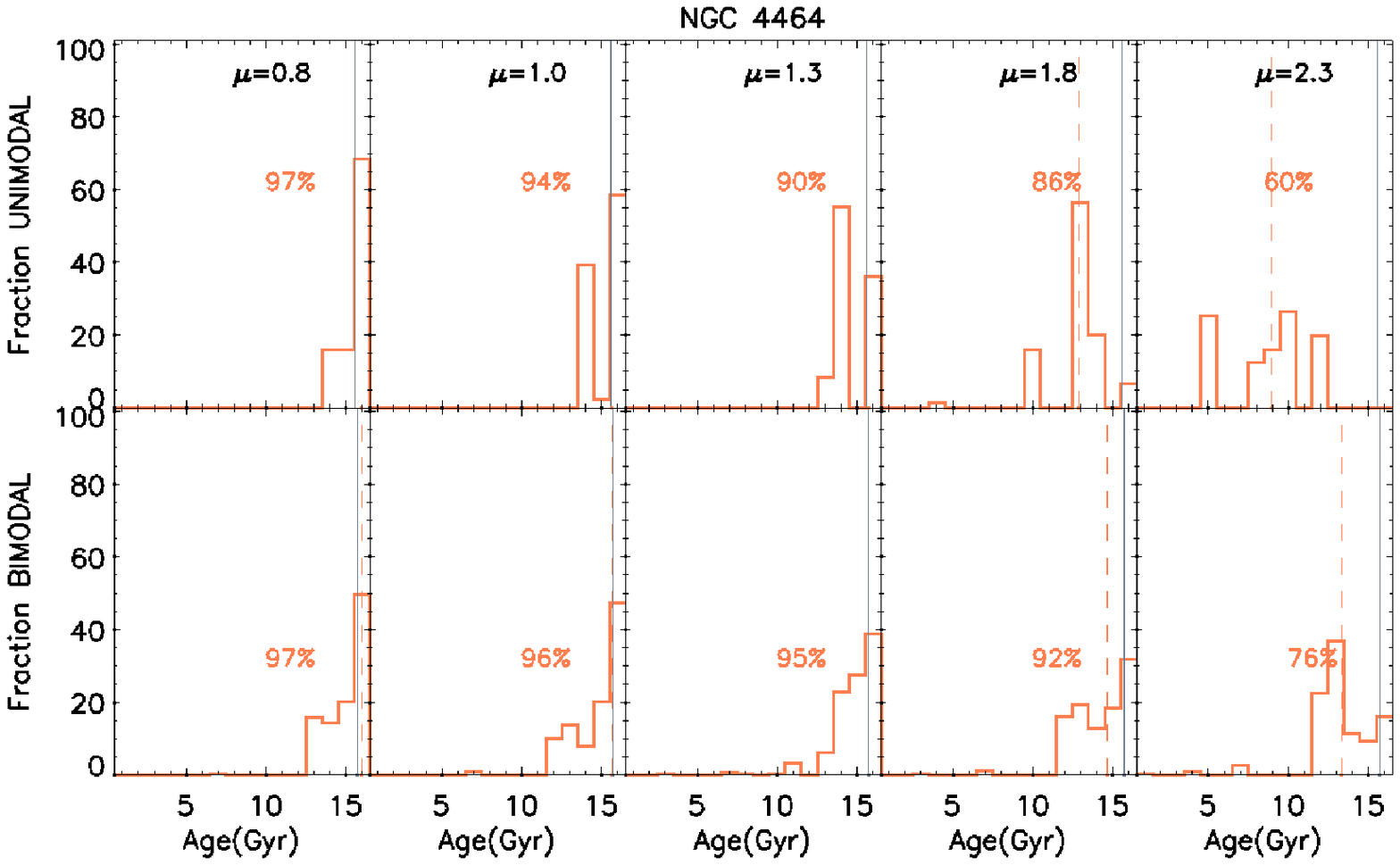}\\ 
        \includegraphics[scale=0.50]{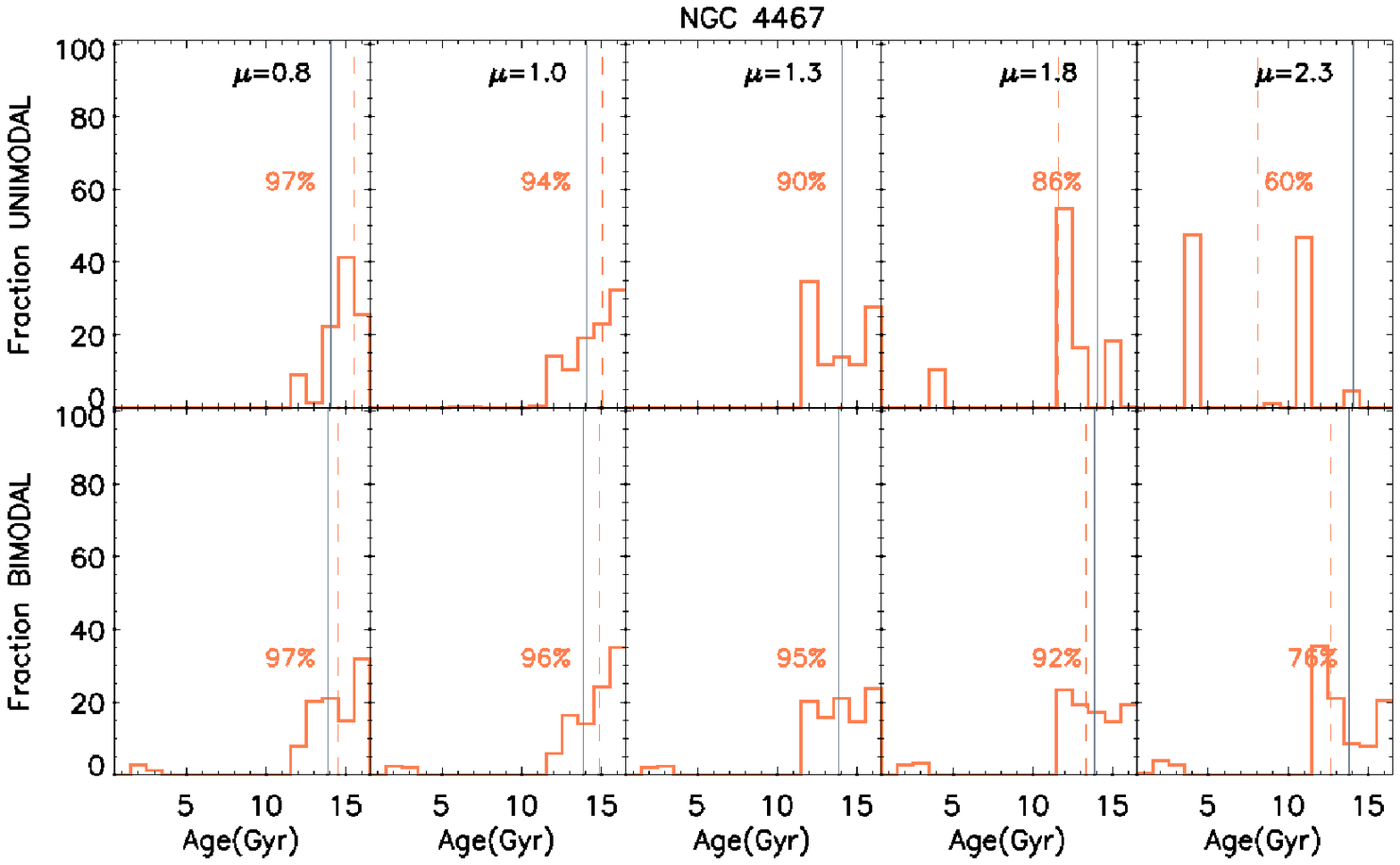}
        \includegraphics[scale=0.50]{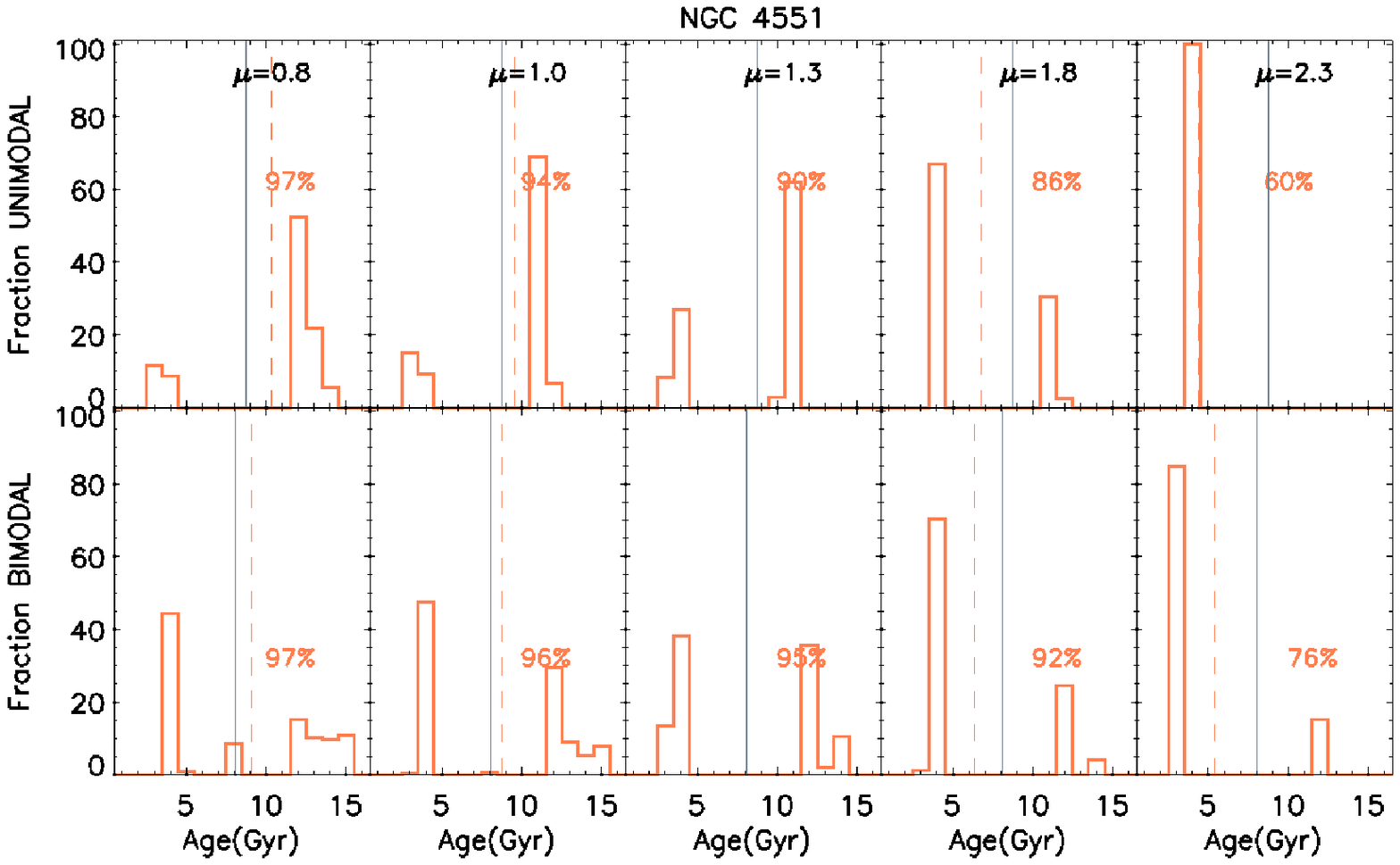}\\ 
\caption{Fig. B1 continued} 
\end{figure*}

\begin{table*}
\centering
\caption{Variations on the stellar population parameters when adopting a UNIMODAL IMF}           
\label{table:B1}     
\begin{tabular}{c|ccccc|ccccc|ccccc}   
\hline\hline      
 & \multicolumn{5}{|c}{Age (Gyr)} & \multicolumn{5}{|c}{$\%$old} & \multicolumn{5}{|c}{Mass ($\times$\,10$^{11}M_{\sun}$)}\\ 
\hline
 Galaxy ID & \multicolumn{5}{|c}{IMF Slope} & \multicolumn{5}{|c}{IMF slope} & \multicolumn{5}{|c}{IMF slope}\\
 &0.8&1.0&1.3&1.8&2.3& 0.8&1.0&1.3&1.8&2.3& 0.8&1.0&1.3&1.8&2.3\\ 
\hline                                                                       
 NYU\,54829  &6.9 &6.6 &5.8 &4.5 &3.2&68 &63 &45 &31 &15 &0.91&1.13&1.31&2.34&4.80\\   
 NYU\,321479 &15.5&14.2&13.7&11.9&7.8&90 &89 &84 &76 &65 &1.41&2.29&2.35&3.74&8.89\\    
 NYU\,685469 &9.7 &9.8 &8.6 &5.6 &3.4&60 &55 &50 &30 &20 &1.03&0.86&1.01&1.76&3.62\\  
 NYU\,796740 &16.2&15.3&14.1&10.1&6.3&95 &91 &89 &86 &61 &2.04&2.71&2.36&4.93&8.73\\    
 NYU\,890167 &9.1 &10.3&10.0&6.6 &3.9&62 &63 &55 &41 &18 &0.31&0.43&0.42&0.86&1.84\\   
 NYU\,896687 &15.8&14.7&14.2&12.6&9.1&93 &91 &86 &78 &66 &2.65&1.79&2.07&4.11&8.50\\
 NYU\,2434587&10.0&8.3 &8.0 &5.2 &3.5&76 &74 &61 &38 &0  &1.37&1.80&1.96&3.30&6.66\\
\hline
 NGC\,4472   &11.4&11.3&11.0&10.2&8.3&99 &99 &97 &93 &86 &1.08&1.01&1.10&2.33&3.02\\
 NGC\,4365   &12.3&12.2&11.6&10.6&8.6&100&100&100&89 &61 &1.33&1.20&1.25&2.02&2.71\\
\hline                                                   
 NGC\,2329   &7.9 &8.4 &8.0 &7.7 &5.8&94 &96 &97 &88 &60 &1.33&1.68&1.91&3.28&5.60\\
 NGC\,4473   &11.5&11.2&10.9&9.4 &7.5&100&100&100&88 &60 &1.85&1.32&1.65&2.81&5.10\\ 
 NGC\,4621   &11.9&11.7&11.2&11.2&9.7&100&100&100&96 &95 &0.33&0.38&0.44&0.56&0.99\\ 
 NGC\,4697   &9.9 &9.7 &8.6 &6.7 &5.0&82 &78 &60 &35 &8  &1.22&1.53&1.92&3.39&5.79\\
 NGC\,5812   &13.1&12.8&12.2&11.5&10.&100&100&100&97 &95 &1.59&0.95&1.62&2.71&5.08\\ 
\hline  
 NGC\,4239   &10.7&9.2 &6.1 &4.3 &3.1&92 &94 &91 &22 &0  &0.06&0.04&0.04&0.09&0.16\\
 NGC\,4339   &10.6&10.2&9.3 &7.4 &5.7&84 &89 &71 &43 &19 &0.23&0.18&0.29&0.76&1.46\\
 NGC\,4387   &13.7&12.9&11.7&9.8 &7.1&100&100&100&84 &35 &0.04&0.04&0.04&0.06&0.09\\ 
 NGC\,4458   &15.6&15.6&14.1&11.1&7.2&97 &94 &90 &86 &60 &0.07&0.06&0.07&0.09&0.15\\ 
 NGC\,4464   &16.8&16.2&15.3&12.7&8.8&100&100&100&98 &75 &0.13&0.11&0.12&0.23&0.39\\
 NGC\,4467   &15.5&15.0&14.1&11.6&8.1&100&100&100&90 &53 &0.06&0.05&0.06&0.08&0.11\\ 
 NGC\,4489   &8.5 &6.4 &4.4 &3.6 &3.0&65 &35 &5  &0  &0  &0.09&0.07&0.08&0.15&0.33\\  
 NGC\,4551   &10.3&9.5 &8.7 &6.7 &4.4&85 &88 &65 &35 &0  &0.22&0.18&0.15&0.31&0.56\\  
\hline                                  
\end{tabular}\\
{Mass-weighted age, fraction of old stellar populations ($>$5\,Gyr) and stellar mass obtained from the derived SFH with a unimodal IMF shape.}
\end{table*}

\begin{table*}
\centering
\caption{Variations on the stellar population parameters when adopting a BIMODAL IMF}           
\label{table:B2}     
\begin{tabular}{c|ccccc|ccccc|ccccc}   
\hline\hline      
 & \multicolumn{5}{|c}{Age (Gyr)} & \multicolumn{5}{|c}{$\%$old} & \multicolumn{5}{|c}{Mass ($\times$\,10$^{11}M_{\sun}$)}\\ 
\hline
 Galaxy ID & \multicolumn{5}{|c}{IMF Slope} & \multicolumn{5}{|c}{IMF slope} & \multicolumn{5}{|c}{IMF slope}\\
 &0.8&1.0&1.3&1.8&2.3& 0.8&1.0&1.3&1.8&2.3& 0.8&1.0&1.3&1.8&2.3\\ 
\hline  
 NYU\,54829  &6.3 &7.5 &6.5 &4.8 &3.4 &60 &62 &52 &38 &14 &0.94&0.78&0.79&0.98&1.34\\   
 NYU\,321479 &12.8&13.8&13.4&11.8&10.4&85 &85 &84 &78 &71 &1.78&1.61&1.62&1.90&2.52\\    
 NYU\,685469 &1.9 &2.1 &1.9 &1.7 &1.5 &0  &0  &0  &0  &0  &0.45&0.42&0.43&0.72&0.92\\  
 NYU\,796740 &15.9&15.9&15.3&15.0&12.7&94 &93 &91 &89 &83 &1.83&2.24&2.11&2.60&2.88\\    
 NYU\,890167 &3.3 &4.1 &4.3 &3.1 &2.8 &25 &25 &27 &28 &13 &0.16&0.15&0.16&0.22&0.45\\   
 NYU\,896687 &14.8&14.7&13.6&12.9&12.8&80 &85 &77 &76 &71 &1.62&1.82&1.51&1.99&2.78\\
 NYU\,2434587&12.4&11.1&10.5&8.1 &7.3 &76 &68 &66 &61 &58 &1.72&1.42&1.38&1.58&2.13\\
\hline                                                               
 NGC\,4472   &11.4&11.5&11.5&11.1&10.6&100&100&98 &95 &88 &2.97&2.79&2.67&2.81&2.97\\
 NGC\,4365   &13.6&13.6&13.0&12.8&12.3&100&100&100&100&100&3.38&2.73&2.40&2.64&3.25\\
\hline 
 NGC\,2329   &7.3 &7.1 &7.1 &5.7 &5.3 &75 &66 &62 &44 &21 &1.82&1.53&1.37&1.55&1.95\\
 NGC\,4473   &11.7&11.5&11.1&10.6&9.8 &100&100&98 &93 &88 &4.47&3.48&2.97&3.28&4.22\\ 
 NGC\,4621   &13.1&12.5&12.3&12.0&11.8&100&100&100&100&98 &0.31&0.24&0.22&0.25&0.13\\ 
 NGC\,4697   &9.7 &9.6 &9.4 &8.8 &8.0 &80 &75 &73 &67 &54 &3.45&2.81&2.65&3.04&3.06\\
 NGC\,5812   &11.1&11.1&11.1&11.0&10.9&100&98 &95 &93 &93 &2.75&2.56&1.98&2.19&2.62\\ 
\hline  
 NGC\,4239   &12.3&12.6&12.2&11.6&9.9 &82 &82 &78 &74 &65 &0.07&0.04&0.04&0.07&0.14\\
 NGC\,4339   &9.9 &10.0&9.2 &8.5 &7.8 &82 &83 &73 &61& 52 &0.53&0.18&0.30&0.76&1.46\\
 NGC\,4387   &13.8&13.5&12.9&11.8&10.1&96 &98 &95 &84 &61 &0.03&0.03&0.03&0.05&0.08\\ 
 NGC\,4458   &16.2&16.1&16.1&15.1&12.5&97 &96 &95 &92 &76 &0.07&0.06&0.07&0.10&0.16\\ 
 NGC\,4464   &16.0&15.6&15.7&14.6&13.3&100&100&100&99 &99 &0.20&0.19&0.19&0.28&0.41\\
 NGC\,4467   &14.5&14.8&13.8&13.3&12.7&96 &95 &95 &93 &92 &0.06&0.06&0.06&0.09&0.11\\ 
 NGC\,4489   &9.7 &10.0&8.9 &7.7 &5.6 &72 &65 &65 &50 &32 &0.09&0.08&0.08&0.15&0.33\\  
 NGC\,4551   &9.0 &8.7 &8.0 &6.4 &5.3 &55 &52 &45 &28 &15 &0.25&0.22&0.15&0.31&0.66\\  
\hline                                  
\end{tabular}\\
{Mass-weighted age, fraction of old stellar populations ($>$5\,Gyr) and stellar mass obtained from the derived SFH with a bimodal IMF shape.}
\end{table*}

\label{lastpage}
\end{document}